\documentclass[conference]{IEEEtran}
\IEEEoverridecommandlockouts
\usepackage{cite}
\usepackage{amsmath,amssymb,amsfonts}
\usepackage{graphicx,color,overpic,psfrag}
\usepackage{xcolor}

\usepackage{graphicx}
\usepackage{textcomp}
\usepackage{fancyhdr}
\usepackage{xcolor}
\usepackage{graphicx}
\usepackage{bm}
\usepackage{subfigure}
\usepackage{multirow}
\usepackage{booktabs}
\usepackage{caption}
\usepackage{booktabs}
\usepackage{amsmath}
\usepackage{algorithm}
\usepackage{amsfonts}
\usepackage{hyperref}
\usepackage{xcolor}
\usepackage{tabularx}
\usepackage{algpseudocode}

\hypersetup{
  colorlinks=true,
  linkcolor=blue,   
  citecolor=blue,   
  urlcolor=blue,    
  linkbordercolor={0 0 1},
  pdfborderstyle={/S/U/W 1}
}

\renewcommand{\cite}[1]{\textcolor{blue}{[\citeonline{#1}]}}

\def\BibTeX{{\rm B\kern-.05em{\sc i\kern-.025em b}\kern-.08em
    T\kern-.1667em\lower.7ex\hbox{E}\kern-.125emX}}

\newcommand{\bl}[1]{\color{black}#1}

\begin{document}

\title{XL-ChannelDiff: An Efficient Diffusion-Based Multi-Domain Near-Field Channel Extrapolation Framework for XL-MIMO Systems
\\

\author{Mengyuan Li, \textit{Graduate Student Member, IEEE}, Yu Han, \textit{Member, IEEE}, Hao Xu, \textit{Senior Member, IEEE},\\ Yongxu Zhu, \textit{Senior Member, IEEE}, Chao-Kai Wen, \textit{Fellow, IEEE}, and Shi Jin, \textit{Fellow, IEEE}\\
}
\thanks{This work was supported by the Mobile Information Networks-National Science and Technology Major Project of China under Grant 2025ZD1301800.\textit{ (Corresponding authors: Yu Han; Shi Jin.)}

M. Li, Y. Han, H. Xu, Y. Zhu, and S. Jin are with the School of Information Science and Engineering, Southeast University, Nanjing 210096, China (e-mail: \{mengyuan\_li, hanyu, hao.xu, yongxu.zhu, jinshi\}@seu.edu.cn). 

C.-K. Wen is with the Institute of Communications Engineering, National Sun Yat-sen University, Kaohsiung 80424, Taiwan (e-mail: chaokai.wen@mail.nsysu.edu.tw).}

}
\maketitle

\begin{abstract}
 Accurate channel state information (CSI) acquisition is essential for unleashing the performance gains of extremely large-scale multiple-input multiple-output (XL-MIMO) systems. However, in near-field regions, CSI acquisition is much more challenging than in the far field due to the high-dimensional channel representation and spherical wavefront propagation. To address this, in this paper, we propose an efficient multi-domain near-field channel extrapolation framework for XL-MIMO systems. Leveraging the conditional denoising diffusion implicit model (CDDIM), our approach enables accurate channel extrapolation across the antenna, frequency, and spatial domains. Specifically, we design a physics-aware CDDIM backbone that incorporates position-embedded patch tokenization and a mask-guided multi-head attention mechanism, enabling the model to exploit position-dependent channel correlations induced by near-field spherical-wave propagation. To ensure high-fidelity extrapolation, we incorporate a Wasserstein GAN (WGAN) discriminator that provides adversarial supervision to the CDDIM during both the training and reverse sampling phases. Additionally, a RePaint-style refinement scheme is introduced to optimize the sampling trajectory, further boosting extrapolation accuracy. Extensive experiments demonstrate the superiority of the proposed framework, achieving superior extrapolation accuracy and robust generalization across diverse domains, varied configurations, and severe masking conditions. 
Codes are available at \url{https://github.com/Lmyxxn/XL-ChannelDiff}.
\end{abstract}


\section{Introduction}
\IEEEPARstart{E}{xtremely} large-scale multiple-input multiple-output (XL-MIMO) has emerged as a key technology for enabling the critical capabilities envisioned in the sixth-generation (6G) systems, including terabit-per-second data rates, centimeter-level sensing and positioning, and massive spatial multiplexing for hundreds of users~\cite{wang2024xlmimo}. In contrast to traditional far-field channel assumptions, near-field channel propagation in XL-MIMO introduces spherical wavefronts~\cite{liu2025nearfield,lu2024tutorial,chen2025nearfield}. Therefore, the near-field channels are inherently higher-dimensional and structurally more complex than conventional far-field channels in massive MIMO systems.
Acquiring accurate channel state information (CSI) is crucial for unleashing the potential of XL-MIMO in high-capacity communication, such as precise localization, and intelligent beamforming~\cite{li2025keypoint}. However, as the number of antennas grows, channel estimation incurs prohibitive overhead in terms of training time, feedback, and computation. These challenges motivate channel extrapolation techniques, which aim to reconstruct complete CSI from partial observations. Rather than estimating all channels directly, these methods exploit domain-specific correlations to infer missing entries, significantly reducing pilot and computational cost~\cite{choi2021experimental,zhang2023ai}.

\subsection{Prior Works}
Research on channel extrapolation has evolved from classical model-driven frameworks to sophisticated data-driven paradigms. Early model-based works, such as parametric channel-based methods, are predicated on the assumption that channel characteristics can be fully encapsulated by a specific set of physical parameters~\cite{han2021multidomain,xu2022sparse,shi2023channel}. Under this premise, extrapolation across time, frequency, and spatial domains is achieved by estimating the variations in these parameters. However, the efficacy of these approaches is intrinsically limited by the gap between idealized channel models and the stochastic nature of practical wireless environments.

Propelled by the remarkable success of deep learning in modeling complex distributions, data-driven approaches have emerged as a compelling alternative to address these limitations. Unlike traditional model-based schemes, these learning-based methods exhibit superior flexibility and robust generalization across diverse propagation environments. Among various deep learning architectures, generative adversarial networks (GANs)~\cite{goodfellow2014gan} have shown strong potential in learning complex data distributions, making them attractive for CSI generation and extrapolation tasks. Their ability to synthesize high-dimensional samples from noise enables data-driven modeling of wireless channels, especially when measurement data is limited. Balevi~\emph{et al.}~\cite{balevi2021wcgan} employed a Wasserstein GAN (WGAN)~\cite{arjovsky2017wgan} to learn high-dimensional channel priors and accurately recover CSI under low pilot conditions. In~\cite{safari2020ul2dl}, a GAN-based image-to-image translation framework was introduced to infer downlink CSI from uplink blocks by treating time-frequency CSI as images, effectively reducing feedback overhead.  Xiao~\emph{et al.}~\cite{xiao2022channelgan} developed ChannelGAN to synthesize statistically realistic CSI samples, reducing reliance on costly measurements. However, the application of GAN-based methods in XL-MIMO scenarios is hindered by significant limitations. Critical issues such as mode collapse severely compromise the diversity of generated CSI, while training instability is further exacerbated by the high dimensionality of the XL-MIMO channels.

Denoising diffusion probabilistic models (DDPMs)~\cite{ho2020ddpm} have emerged as another powerful class of generative frameworks that learn to reverse a stochastic Markovian noising process. By incrementally perturbing clean data with Gaussian noise during the forward phase, DDPMs learn to reconstruct high-fidelity samples from complex distributions through a sequence of iterative denoising steps. Following their success in computer vision and audio synthesis~\cite{rombach2022ldm, eschweiler2024microscopy, popov2021gradtts}, DDPMs have recently gained significant momentum in the wireless communication domain. They have demonstrated superior capability in capturing the intricate statistical characteristics of wireless channels, yielding promising performance in tasks such as channel estimation~\cite{wu2023cddm, yi2024cediff, zhang2023scorebased,fesl2024diffusion,zhou2025generative}, signal detection~\cite{signal_ddpm}, and multi-domain channel extrapolation~\cite{zhang2025cddpm, lee2024cddim}. 
For CSI extrapolation, DDPMs provide a principled framework to recover the full channel from partial observations by formulating the problem as a conditional generation task. Specifically, a conditional DDPM (CDDPM)-based channel acquisition scheme was proposed in~\cite{zhang2025cddpm} for reconfigurable intelligent surface-assisted systems, which integrates decision Transformers with diffusion priors to ensure temporal consistency under extreme pilot sparsity. Similarly, a conditional denoising diffusion implicit model (CDDIM)-based channel generation framework was proposed in~\cite{lee2024cddim}. 
By conditioning the denoising diffusion implicit model on user positions, this framework generates high-fidelity synthetic channel samples for data augmentation, thereby mitigating the scarcity of measured channel data.
Moreover, a comparative analysis between WGAN and DDPM was presented in~\cite{sengupta2023diffusion}, demonstrating that diffusion models offer superior distributional fidelity and sample diversity for true channel augmentation.

{\bl
Despite these advances, diffusion-based channel extrapolation for XL-MIMO systems remains challenging. First, the Markovian reverse sampling process of DDPMs usually requires a large number of denoising iterations, resulting in considerable inference latency. Second, existing studies mainly consider far-field or low-dimensional MIMO scenarios, and their network designs usually do not explicitly account for the near-field characteristics of XL-MIMO channels. Due to the high dimensionality and spherical-wave characteristics of near-field channels in XL-MIMO systems, learning the conditional posterior for channel extrapolation is non-trivial. Third, most existing diffusion frameworks are tailored to a single extrapolation domain, whereas channel extrapolation in different domains involves different channel correlation patterns. 

\subsection{Main Contributions}
To address these challenges, we summarize our main contributions in this paper as follows:
\begin{itemize}
   
    \item \textbf{Physics-Aware CDDIM Backbone:} 
    We first analyze spherical-wave-induced channel correlations across the antenna, frequency, and spatial domains to guide the backbone design. 
    The 2D/3D convolutional patch embeddings and positional encodings (PEs) provide physics-aware tokenization. 
    The Transformer blocks then learn dependencies among these physically ordered tokens, while mask-guided attention and a tailored RePaint scheme exploit reliable observations and enforce consistency during generation.

    \item \textbf{Multi-Domain Channel Extrapolation Support:} To address the constraints of domain-specific designs, we introduce a modular architecture that enables a Transformer backbone to process tokenized inputs from antenna, frequency, and spatial domains seamlessly without requiring structural changes. The proposed framework can achieve consistently superior extrapolation performance across these diverse domains.

    \item \textbf{WGAN-Enhanced CDDIM Channel Extrapolator:} We incorporate adversarial supervision into both training and reverse sampling pipelines of the proposed CDDIM backbone. Specifically, a WGAN-based discriminator guides the designed CDDIM learning process to enhance the distributional fidelity and diversity of extrapolated channels. This synergy facilitates accurate channel extrapolation with significantly fewer sampling steps and improved robustness against mask sparsity.

     \item \textbf{Extensive Experiments on Performance Evaluation:} We conduct comprehensive evaluations across the antenna, frequency, and spatial domains under a wide range of mask ratios and different masking patterns. Beyond benchmarking against competitive baselines, we evaluate the generalization ability across diverse carrier frequencies, array sizes, and both near-field and far-field regimes. The results demonstrate the proposed framework's superior performance and robust adaptability to diverse system configurations.
\end{itemize}
}

{\bf Organization.}
The remainder of this paper is organized as follows. Section~\ref{sec:Channel Model} introduces the near-field channel model and correlation analysis. Section~\ref{sec:Multi-Domain Channel Extrapolation} details the proposed WGAN-enhanced CDDIM channel extrapolation framework. Section~\ref{sec:experiments} presents the experimental results and analysis. Finally, Section~\ref{sec:conclusion} concludes the paper.

\textbf{Notations.} Scalars are denoted by regular letters, while vectors and matrices are denoted by lowercase and uppercase boldface letters, respectively. $\mathcal{CN}$ denotes the complex Gaussian distribution. The expectation operator is denoted by $\mathbb{E}\{\cdot\}$. The $\ell_2$ norm of a vector $\mathbf{x}$ is denoted by $\|\mathbf{x}\|_2$. The Frobenius norm of a complex matrix $\mathbf{A}$ is denoted by $\|\mathbf{A}\|_F$, and the inner product is defined as $\langle \mathbf{A},\mathbf{B}\rangle_F=\mathrm{tr}(\mathbf{A}^{\mathrm H}\mathbf{B})$. For a complex scalar, vector, or matrix, $(\cdot)^*$, $(\cdot)^{\mathrm T}$, and $(\cdot)^{\mathrm H}$ denote complex conjugation, transpose, and Hermitian transpose, respectively. The operator $\Re\{\cdot\}$ denotes the real part of a complex quantity. The modulus of a complex number $z$ is denoted by $|z|$. $\odot$ denotes the Hadamard product. $\mathbf{I}$ denotes the identity matrix, and $\nabla_{\cdot}\mathcal{F}(\cdot)$ represents the gradient of a function $\mathcal{F}(\cdot)$.


\section{Channel Correlation Model}
\label{sec:Channel Model}

In this section, we first introduce the near-field channel model and then analyze the antenna-, frequency-, and spatial-domain channel correlations, revealing how antenna displacement, frequency separation, and user-position variation lead to correlation decay behaviors. 
These correlation characteristics provide the theoretical basis for the proposed multi-domain extrapolation framework.

\subsection{Near-Field Channel Model}

We consider a near-field XL-MIMO system, where a single-antenna user equipment (UE) communicates with a base station (BS) equipped with a uniform planar array (UPA). 
Let $\mathbf{u}\in\mathbb{R}^3$ denote the UE location and $\mathbf{r}_m\in\mathbb{R}^3$ denote the position of the $m$-th BS antenna element.
To characterize the multi-domain correlations in near-field XL-MIMO channels, we use the following geometry-based multipath representation with line-of-sight (LoS) and equivalent last-bounce non-line-of-sight (NLoS) components.

At frequency $f$, the channel observed at the $m$-th antenna element is modeled as
\begin{equation}
\label{eq:channel_response}
h_m(f,\mathbf{u}) 
= 
\sum_{l=1}^{L}
\alpha_l
\exp\left(
-j\frac{2\pi f}{c}d_{l,m}(\mathbf{u})
\right),
\end{equation}
where $\alpha_l$ denotes the complex gain of the $l$-th path, $c$ is the speed of light, and $d_{l,m}(\mathbf{u})$ is the element-dependent propagation distance. 
For the LoS component, i.e., $l=1$, the propagation distance between UE and the $m$-th antenna is given by
\begin{equation}
\label{eq:los_distance}
d_{1,m}(\mathbf{u})
=
\|\mathbf{u}-\mathbf{r}_m\|_2 .
\end{equation}
For the $l$-th NLoS component with $l>1$, we use the equivalent last-bounce representation as
\begin{equation}
\label{eq:nlos_distance}
d_{l,m}(\mathbf{u})
=
d_l^{\mathrm{pre}}(\mathbf{u})
+
\|\mathbf{s}_l-\mathbf{r}_m\|_2,
\end{equation}
where $\mathbf{s}_l$ denotes the equivalent last-bounce scatterer, and $d_l^{\mathrm{pre}}(\mathbf{u})$ is the accumulated propagation distance before this scatterer. 
For the simplified single-bounce case, we have
\begin{equation}
\label{eq:single_bounce_distance}
d_l^{\mathrm{pre}}(\mathbf{u})
=
\|\mathbf{u}-\mathbf{s}_l\|_2 .
\end{equation}

Different from the far-field plane-wave assumption, the distance terms in \eqref{eq:los_distance} and \eqref{eq:nlos_distance} depend on the exact antenna-element position $\mathbf{r}_m$. 
Therefore, the propagation phase varies across the XL-MIMO aperture, leading to spherical-wavefront characteristics and spatially non-stationary correlations.

\subsection{Correlation Analysis}

For notational brevity, we denote 
$h_x \triangleq h_{m_x}(f_x,\mathbf{u}_x)$ and 
$h_y \triangleq h_{m_y}(f_y,\mathbf{u}_y)$.
The normalized channel correlation coefficient is defined as
\begin{equation} 
\label{eq:corr_def}
\mathrm{R}(h_x,h_y)
= \frac{\mathbb{E}[h_x h_y^*]}
       {\sqrt{\mathbb{E}[|h_x|^2]\,\mathbb{E}[|h_y|^2]}} .
\end{equation}
For theoretical tractability, we adopt the uncorrelated scattering assumption, i.e.,
$\mathbb{E}[\alpha_l\alpha_{l'}^*]=0$ for $l\neq l'$.
The normalized power fraction of the $l$-th path is defined as
\begin{equation}
\label{eq:rho_effective}
\rho_l
\triangleq
\frac{
\mathbb{E}\!\left[|\alpha_l|^2\right]
}
{
\sum_{k=1}^{L}
\mathbb{E}\!\left[|\alpha_k|^2\right]
}.
\end{equation}

In the \textit{antenna domain}, for antenna elements $m_x$ and $m_y$ at the same $(f,\mathbf{u})$, the correlation can be approximated as
\begin{equation}
\label{eq:corre_antenna}
\mathrm{R}^{(m)}(h_x,h_y) 
\approx 
\sum_{l=1}^{L} 
\rho_l
\exp\left(
-j\frac{2\pi f}{c}
\left[
d_{l,m_x}(\mathbf{u})-d_{l,m_y}(\mathbf{u})
\right]
\right).
\end{equation} 
Thus, antenna-domain correlation is governed by the path-length differences across antenna elements. 
In the near field, these differences depend on the absolute antenna positions rather than only on antenna spacing in the far field, resulting in position-dependent correlations over the large aperture.

In the \textit{frequency domain}, for a fixed antenna element $m$ and user position $\mathbf{u}$, the geometric path length $d_{l,m}(\mathbf{u})$ is frequency-invariant. 
The correlation between two frequencies $f_x$ and $f_y$ can be approximated as
\begin{equation}
\label{eq:corre_fre}
\mathrm{R}^{(f)}(h_x,h_y) 
\approx 
\sum_{l=1}^{L} 
\rho_l
\exp\left(
-j\frac{2\pi (f_x-f_y)}{c}
d_{l,m}(\mathbf{u})
\right).
\end{equation}
Therefore, frequency-domain correlation is determined by the frequency separation and the path delays. 
As $\Delta f=|f_x-f_y|$ increases, the phase rotations of different paths become more dispersed, leading to correlation decay across subcarriers. 
Moreover, since $d_{l,m}(\mathbf{u})$ varies with $m$, the frequency selectivity also becomes antenna-dependent in near-field XL-MIMO systems.

In the \textit{spatial domain}, for two nearby user positions $\mathbf{u}_x$ and $\mathbf{u}_y$ observed at the same $(m,f)$, and assuming that the dominant paths can be approximately associated across the two positions, the correlation can be approximated as
\begin{equation}
\label{eq:corre_spatial}
\mathrm{R}^{(s)}(h_x,h_y) 
\approx 
\sum_{l=1}^{L} 
\rho_l
\exp\left(
-j\frac{2\pi f}{c}
\left[
d_{l,m}(\mathbf{u}_x)-d_{l,m}(\mathbf{u}_y)
\right]
\right).
\end{equation}
Therefore, spatial-domain correlation is mainly determined by the path-length variation caused by the change of user position. 
A larger spatial separation generally introduces larger differences in propagation distances and angles, which increases the phase dispersion among multipath components and reduces the channel correlation.

From \eqref{eq:corre_antenna}--\eqref{eq:corre_spatial}, the multi-domain correlation can be interpreted as a weighted summation of path-dependent phasors. 
The correlation remains high when the path-dependent phase differences are aligned, while it decreases when these phase differences become dispersed. 
Therefore, antenna displacement, frequency separation, and user-position variation produce different phase-mismatch patterns, leading to domain-dependent correlation decay.
{\bl We also provide an empirical visualization of the channel correlation decay under the default simulation setup in Sec.~\ref{sec:experiments}, as shown later in Fig.~\ref{fig:multi_domain_correlation}.}
The above correlations are rooted in the spherical-wavefront nature of near-field propagation. 
Unlike far-field channels dominated mainly by angular separation, near-field steering vectors contain additional quadratic phase terms, as revealed by the Fresnel approximation~\cite{Fresnel}. 
These terms change the correlation behavior and break the translational invariance commonly assumed by conventional estimators, which motivates a physics-aware generative framework for capturing spatially variant multi-domain dependencies.

\section{Proposed WGAN-Enhanced CDDIM Channel Extrapolation Framework}
\label{sec:Multi-Domain Channel Extrapolation}

Based on the channel correlation model, we first present the channel extrapolation problem formulation and the diffusion processes, followed by the CDDIM-based extrapolation algorithm. 
To capture the spatially variant characteristics of near-field channels, we design a physics-aware backbone with position-dependent tokenization. 
Furthermore, a WGAN is integrated to provide adversarial supervision during training and sampling, thereby improving the fidelity and structural consistency of the extrapolated channels.

\begin{figure}[t]
\centering

\begin{minipage}[t]{0.485\linewidth}
    \centering
    \includegraphics[width=\linewidth]{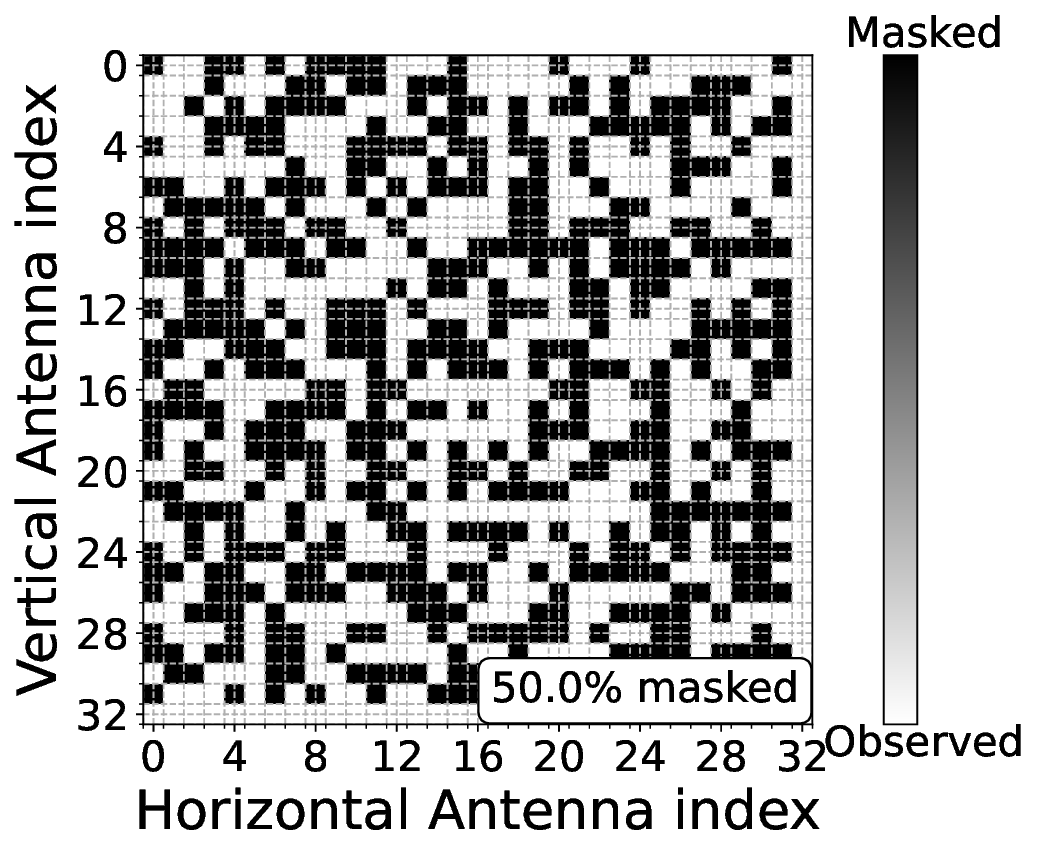}
    \vspace{-0.3em}
    \small (a) Antenna, random
\end{minipage}%
\hspace{0.01\linewidth}%
\begin{minipage}[t]{0.485\linewidth}
    \centering
    \includegraphics[width=\linewidth]{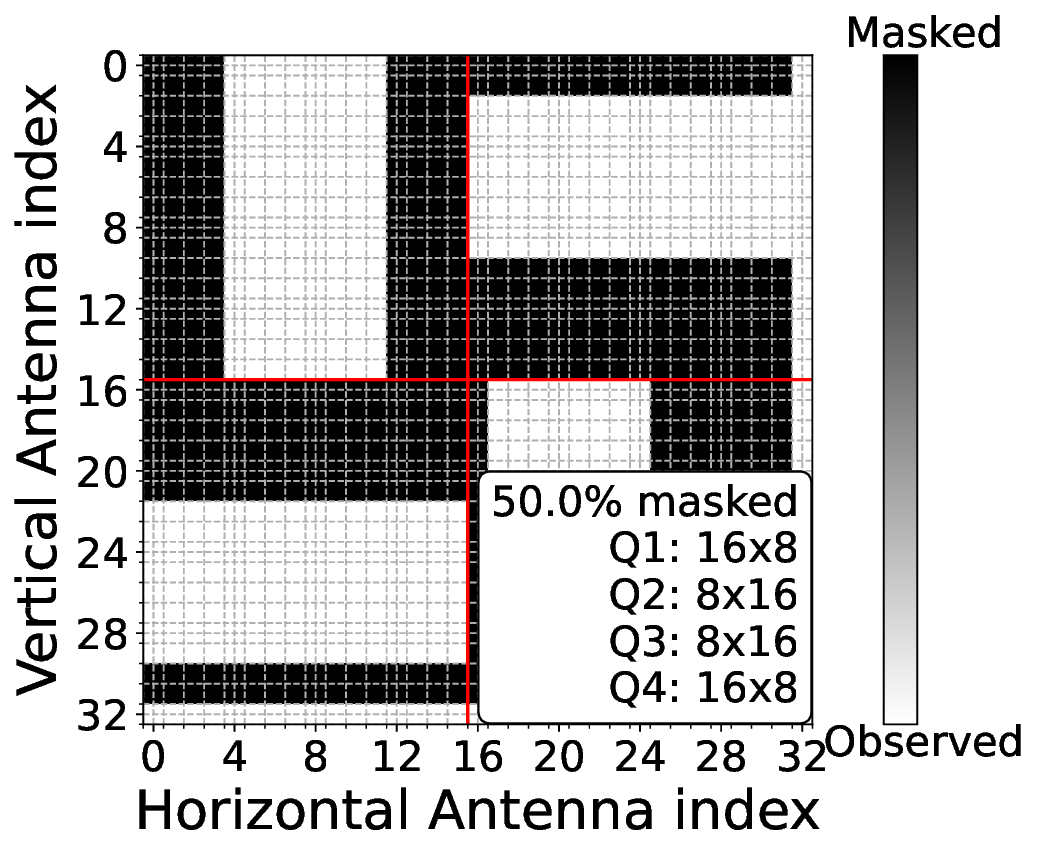}
    \vspace{-0.3em}
    \small (b) Antenna, block-structured
\end{minipage}

\vspace{0.2em} 

\begin{minipage}[t]{0.485\linewidth}
  \centering
  \includegraphics[width=\linewidth]{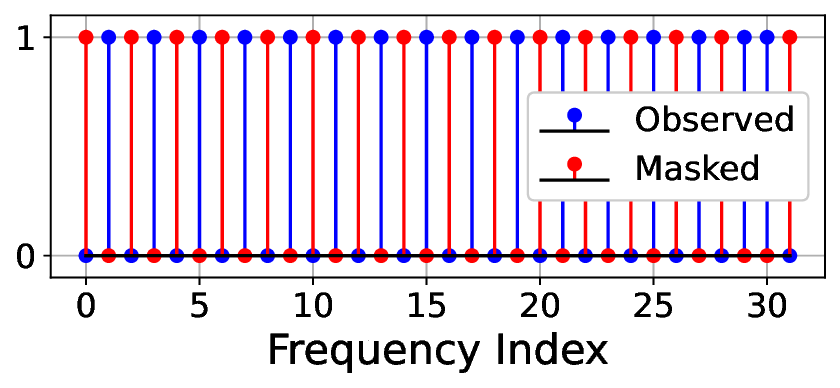}
  \vspace{-0.3em}
  \small (c) Frequency, structured
\end{minipage}%
\hspace{0.01\linewidth}%
\begin{minipage}[t]{0.485\linewidth}
  \centering
  \includegraphics[width=\linewidth]{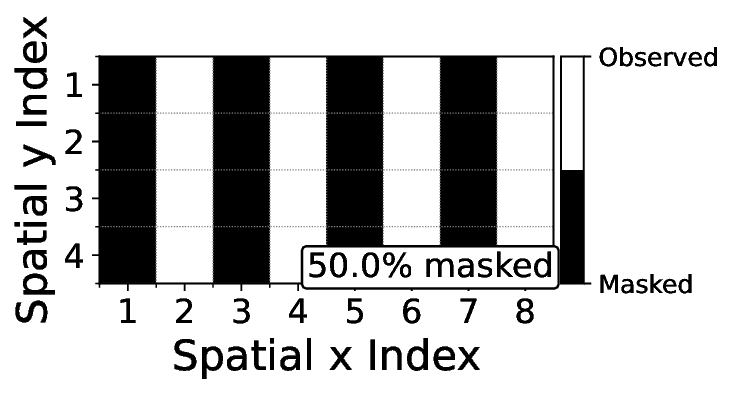}
  \vspace{-0.3em}
  \small (d) Spatial, structured
\end{minipage}

\vspace{0.1em}
\caption{{\bl Examples of mask patterns in the antenna, frequency, and spatial domains for a mask ratio of $\gamma = 0.5$.}}
\label{fig:mask_patterns_combined}
\end{figure}

\begin{figure}[t]
\centering
\subfigure[2D antenna domain extrapolation]{
    \includegraphics[scale=0.14]{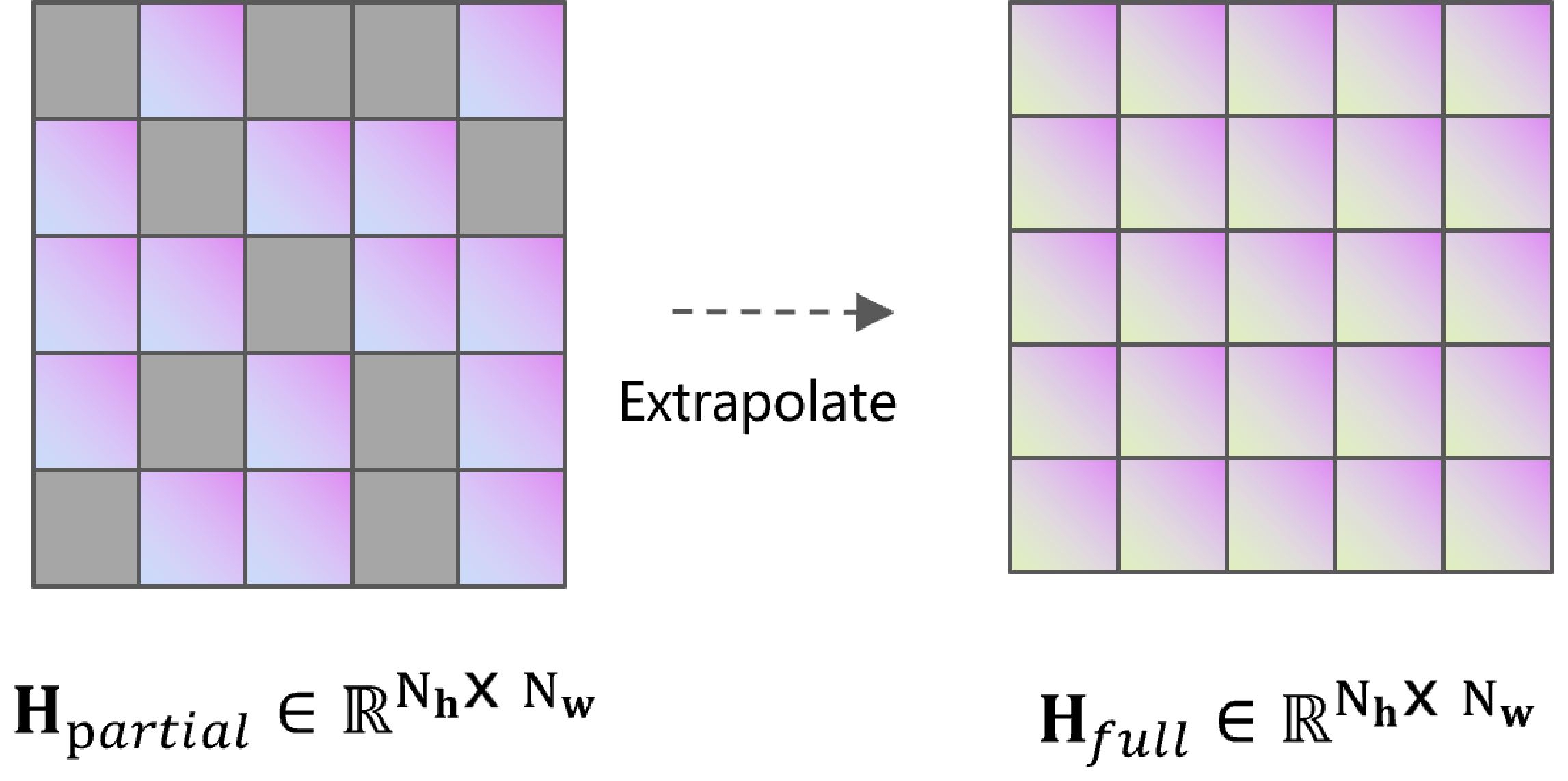}%
}

\vspace{-0.4em} 

\subfigure[3D frequency and spatial-domain extrapolation]{
    \includegraphics[scale=0.13]{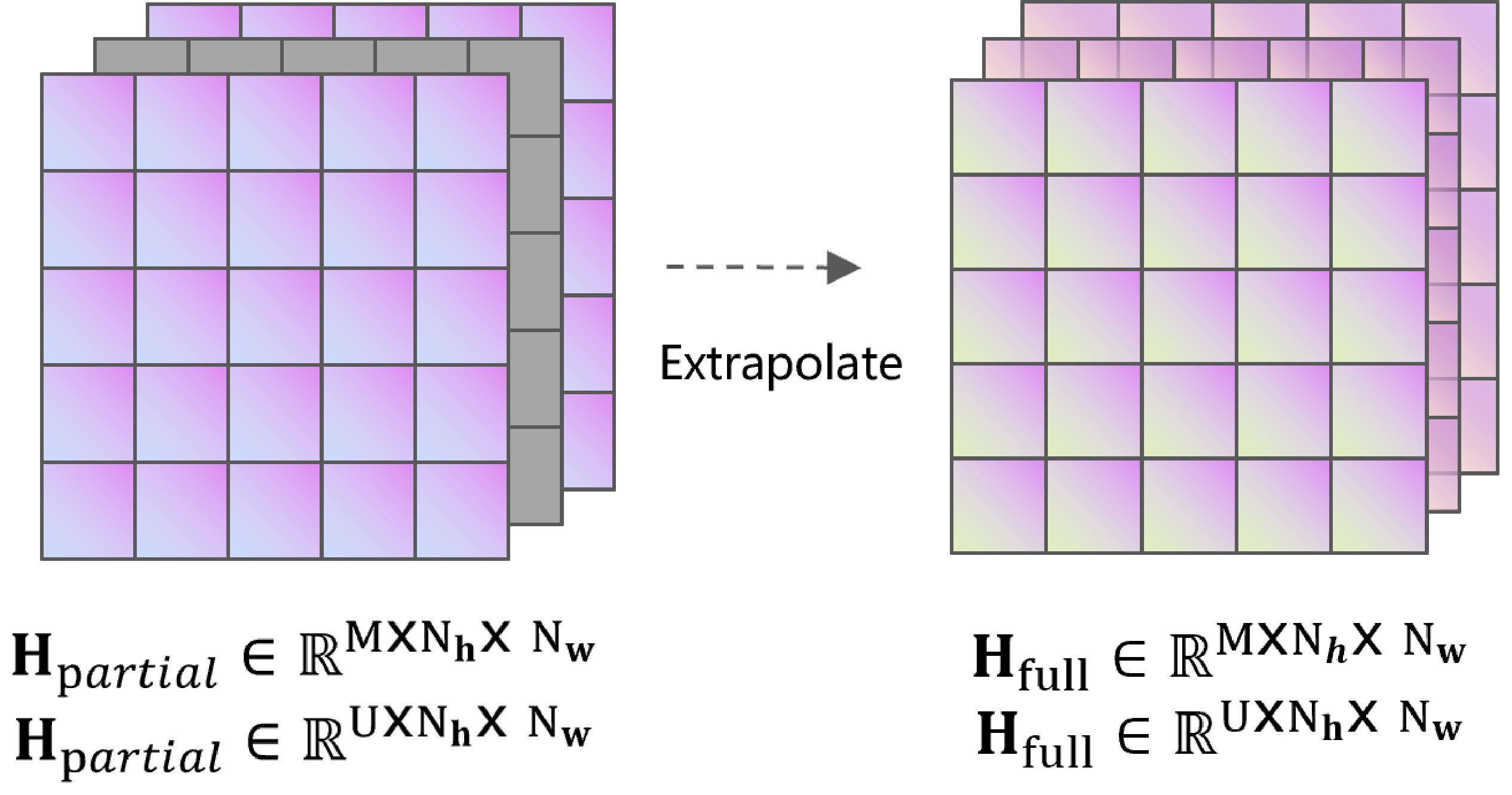}
}

\vspace{-0.4em} 

\caption{Multi-domain channel extrapolation tasks. Gray entries denote unknown channels.}

\vspace{-0.4em} 

\label{fig:channel_extrapolation_tasks}
\end{figure}

\subsection{Multi-Domain Channel Extrapolation Task Formulation}
 We use $\mathbf{H}_{\text{full}}$ to represent the GT full channel, and introduce a binary sampling mask $\mathbf{M}$ to indicate the observed entries, where $1$ represents an observed entry and $0$ denotes a masked entry. Consequently, the available partial channel, denoted as $\mathbf{H}_{\text{partial}}$ is obtained by
\begin{equation}
    \mathbf{H}_{\text{partial}} = \mathbf{M} \odot \mathbf{H}_{\text{full}}.
\end{equation}
The configuration of the sampling mask $\mathbf{M}$, as exemplified in Fig. \ref{fig:mask_patterns_combined}, is tailored to represent distinct physical constraints inherent in each domain. Specifically, for the antenna domain, we consider random masks (Fig. \ref{fig:mask_patterns_combined}(a)) to reflect irregular element failures or asynchronous access of the antenna array, and block-structured masks (Fig. \ref{fig:mask_patterns_combined}(b)) to simulate hardware-constrained architectures where only contiguous subarrays remain active due to cost or power limitations. In the frequency and spatial domains, $\mathbf{M}$ adheres to structured sampling patterns (Figs. \ref{fig:mask_patterns_combined}(c) and \ref{fig:mask_patterns_combined}(d)) to represent the comb-type pilot pattern and spatial sampling, respectively. By characterizing $\mathbf{M}$ through these practical scenarios, the framework is trained to accurately reconstruct $\mathbf{H}_{\text{full}}$ by exploiting latent channel correlations.

As Fig.~\ref{fig:channel_extrapolation_tasks} illustrates, in the \emph{antenna domain}, $\mathbf{H}_{\text{full}}\in\mathbb{C}^{N_h\times N_w}$ collects the CSI over all UPA elements, while $\mathbf{H}_{\text{partial}}\in\mathbb{C}^{N_h\times N_w}$ contains observations on only a subset of antennas, i.e., $\mathbf{M} \in \{0, 1\}^{N_h \times N_w}$ is sparse over the $(N_h,N_w)$ antenna array, where $N_h$ and $N_w$ are the number of antennas along the vertical and horizontal dimensions, respectively. The goal is to recover the missing antenna entries by leveraging inter-antenna correlations characterized in~\eqref{eq:corre_antenna}.
In the \emph{frequency domain}, $\mathbf{H}_{\text{full}}\in\mathbb{C}^{M\times N_h\times N_w}$ represents the CSI over $M$ subcarriers and all antenna elements. Here, $\mathbf{M} \in \{0, 1\}^{M \times N_h \times N_w}$ is sparse along the subcarrier dimension, so that $\mathbf{H}_{\text{partial}}\in\mathbb{C}^{M\times N_h\times N_w}$ contains CSI on only a subset of subcarriers. The objective is to infer the unobserved subcarriers by exploiting frequency-domain correlations as in~\eqref{eq:corre_fre}.
In the \emph{spatial domain}, $\mathbf{H}_{\text{full}}\in\mathbb{C}^{U\times N_h\times N_w}$ stacks the CSI over $U$ spatial sampling points. The mask $\mathbf{M} \in \{0, 1\}^{U \times N_h \times N_w}$ is sparse along the spatial dimension, yielding $\mathbf{H}_{\text{partial}}\in\mathbb{C}^{U\times N_h\times N_w}$ observed at only a subset of spatial points. The task is to extrapolate the CSI to unobserved spatial locations by leveraging spatial-domain correlations described in~\eqref{eq:corre_spatial}.

\subsection{CDDIM-Based Channel Extrapolation Algorithm}
Leveraging the superior generative capabilities of diffusion models in capturing complex high-dimensional data distributions, we draw inspiration from recent advances in conditional diffusion models to develop a channel extrapolation solution tailored for XL-MIMO systems. Specifically, to alleviate the high inference latency inherent in standard Markovian sampling, we adopt the CDDIM algorithm for efficient channel extrapolation. Unlike the traditional Markovian process in DDPM, CDDIM utilizes the non-Markovian diffusion processes that enables a deterministic generative trajectory. This formulation allows for accelerated sampling by skipping steps while preserving the high fidelity of the original training objective~\cite{lee2024cddim}.
Within this framework, we treat the channel extrapolation task as a conditional generation problem, where the observed partial channel $\mathbf{H}_{\text{partial}}$ serves as the condition that guides both the noise prediction network and the iterative reverse updates. By anchoring the generative trajectory to these available measurements, CDDIM ensures that the extrapolated full channel $\mathbf{H}_{\text{full}}$ maintains physical consistency with the partial observations. For diffusion notation, we denote the ground-truth (GT) full channel as $\mathbf H_0 \triangleq \mathbf H_{\text{full}}$. The comprehensive extrapolation pipeline is characterized by the following key processes:

\paragraph{Forward Process}  
The forward process of the proposed framework follows the principles of standard DDPMs, which transforms the channel into a Gaussian prior by incrementally injecting noise. Specifically, the transition between successive time steps is defined as
\begin{equation}
\mathbf{H}_t = \sqrt{1 - \beta_t} \, \mathbf{H}_{t-1} + \sqrt{\beta_t} \, \boldsymbol{\varepsilon}_{t}, 
\end{equation}
where $\boldsymbol{\varepsilon}_{t} \sim \mathcal{CN}(\mathbf{0}, \mathbf{I})$ represents the independent and identically distributed (i.i.d.) standard complex Gaussian noise at time step $t$. The noise variance schedule is controlled by $\beta_t$, which is designed as an increasing sequence $0 < \beta_1 < \beta_2 < \cdots < \beta_{T-1}$.  This schedule ensures that the underlying data distribution gradually converges toward a standard complex Gaussian distribution as $t$ approaches $T$. 
By leveraging the property of Gaussian superposition, the marginal distribution of the noisy channel $\mathbf{H}_t$ conditioned on the GT channel $\mathbf{H}_0$ can be derived as 
\begin{equation}
q(\mathbf{H}_t|\mathbf{H}_0) = \mathcal{CN}\left(\mathbf{H}_t; \sqrt{\bar{\alpha}_t} \mathbf{H}_0, (1-\bar{\alpha}_t)\mathbf{I}\right),
\label{Eq:ddpm}
\end{equation}
where $\alpha_i = 1 - \beta_i$, and $\bar{\alpha}_t = \prod_{i=1}^t \alpha_i$ represents the cumulative product of the noise-scheduling parameters. Consequently, the noisy channel at any arbitrary time step $t$ can be expressed in a closed-form reparameterization as
\begin{equation} 
\mathbf{H}_t=\sqrt{\bar{\alpha}_t}\mathbf{H}_0+\sqrt{1-\bar{\alpha}_t}\,\boldsymbol{\epsilon}_t,
\label{eq:noisy_channel}
\end{equation}
where $\boldsymbol{\epsilon}_t$ is the effective noise of time step $t$.

\paragraph{Noise Prediction Process}
The CDDIM algorithm aims to learn the effective noise $\boldsymbol{\epsilon}_t$ injected at each timestep $t$ through a neural network. During the training phase, as the GT full channel $\mathbf{H}_0$ is accessible, we first generate the noisy sample $\mathbf{H}_t$ by applying the forward perturbation process via~\eqref{eq:noisy_channel}. The transition posterior is analytically derivable using Bayes' rule as follows:
\begin{equation}
\begin{aligned}
q&(\mathbf{H}_{t-1}\mid \mathbf{H}_t,\mathbf{H}_0)
\\[2pt]
&= \frac{q(\mathbf{H}_t\mid \mathbf{H}_{t-1},\mathbf{H}_0)\,q(\mathbf{H}_{t-1}\mid \mathbf{H}_0)}{q(\mathbf{H}_t\mid \mathbf{H}_0)}
\\[2pt]
&=\;\mathcal CN\!\Big(
\mathbf H_{t-1}\,;\;
\underbrace{\tfrac{1}{\sqrt{\alpha_t}}\!\Big(\mathbf{H}_t-\tfrac{1-\alpha_t}{\sqrt{1-\bar{\alpha}_t}}\,
\boldsymbol{\epsilon}_t\Big)}_{\displaystyle \boldsymbol\mu(\mathbf H_t,\mathbf H_0,t)}\!,
\;
\underbrace{\tfrac{\beta_t(1-\bar{\alpha}_{t-1})}{1-\bar{\alpha}_t}}_{\displaystyle \boldsymbol{\Sigma}_t}\mathbf I
\Big),
\end{aligned}
\label{eq:posterior-bayes-gauss}
\end{equation}
where the covariance $\boldsymbol{\Sigma}_t$ is determined by the predefined noise-scheduling hyper-parameters. The detailed derivation of~\eqref{eq:posterior-bayes-gauss} is deferred to Appendix~\ref{app:posterior-derivation}.

\begin{figure*}[t]
    \centering
    \includegraphics[width=0.95\linewidth]{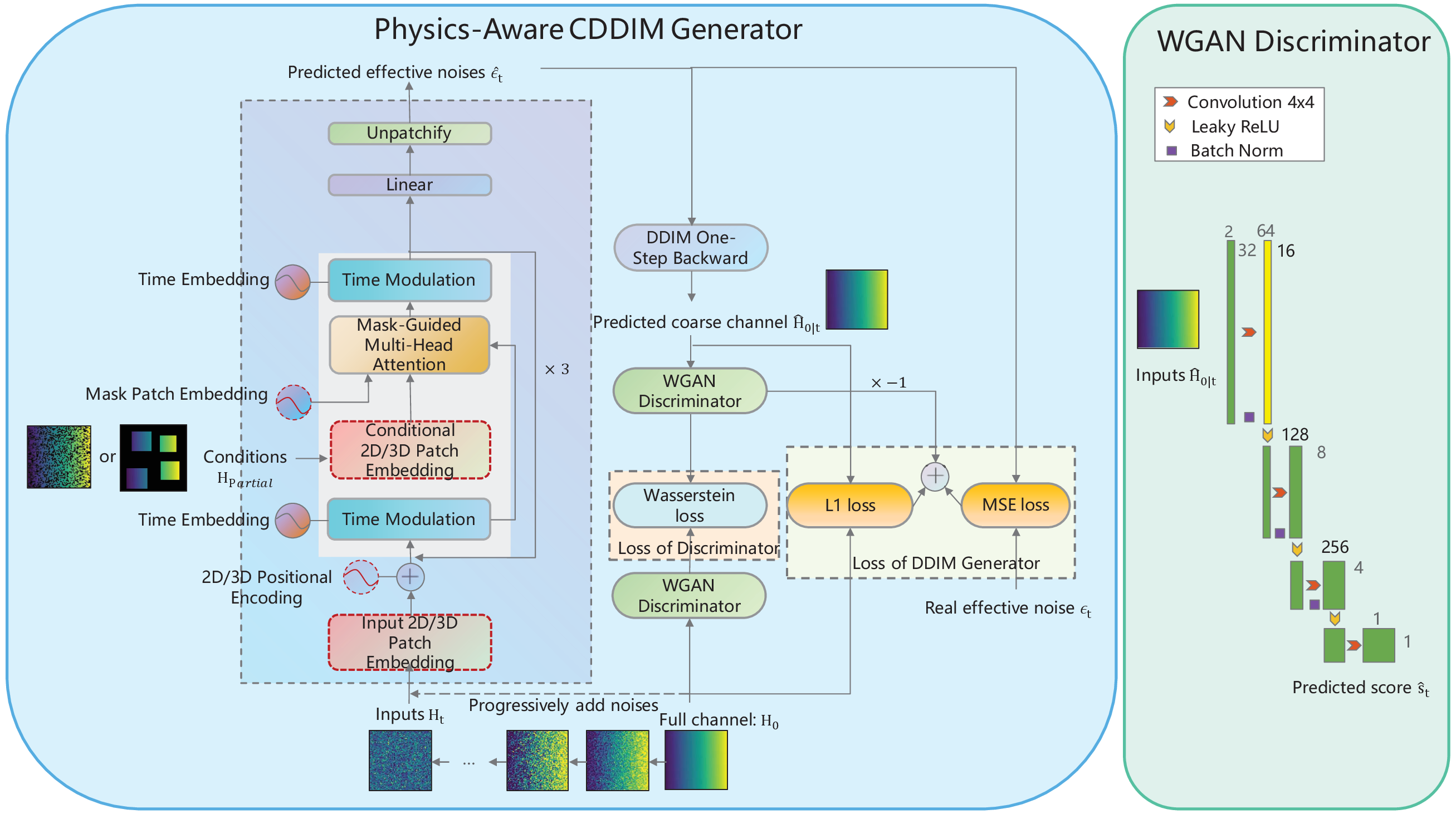}
    \caption{{\bl The architecture of the proposed WGAN-enhanced CDDIM channel extrapolation framework, featuring a physics-aware CDDIM generator and a WGAN discriminator.}}
    \label{fig:diffuser}
\end{figure*}

In the inference process, the clean full channel $\mathbf H_0$ is unavailable. To address this, we approximate the reverse transition with a conditional generative model that leverages the observed partial channel $\mathbf H_{\text{partial}}$ as a guiding prior. The estimated reverse posterior is parameterized with a learnable mean $\boldsymbol{\mu}_{\theta}$ and the schedule-dependent covariance $\boldsymbol{\Sigma}_t$ as:
\begingroup
\small
\begin{equation}
q_{\theta}\!\big(\mathbf H_{t-1}\mid \mathbf H_t,\mathbf H_{\text{partial}},t\big)
= \mathcal{CN}\!\Big(\mathbf H_{t-1}\,;\; \boldsymbol{\mu}_{\theta}(\mathbf H_t,\mathbf H_{\text{partial}},t),\; \boldsymbol{\Sigma}_t\Big),
\label{eq:q-theta}
\end{equation}
\endgroup
where the mean $\boldsymbol{\mu}_{\theta}$ is formulated to account for the noise predicted by the neural network $\hat{\boldsymbol{\epsilon}}_{t}$, and can be expressed as:
\begin{equation}
\boldsymbol{\mu}_{\theta}(\mathbf H_t,\mathbf H_{\text{partial}},t)
=\tfrac{1}{\sqrt{\alpha_t}}\!\left(
\mathbf H_t - \tfrac{1-\alpha_t}{\sqrt{1-\bar\alpha_t}}\,
\hat{\boldsymbol{\epsilon}}_{t}\right),
\label{eq:mu-theta}
\end{equation}
and the estimated noise is predicted by the noise prediction network $\mathcal{G}_{\theta}$ through:
\begin{equation}
  \hat{\boldsymbol{\epsilon}}_{t}
  \;=\;
\mathcal{G}_{\theta}\!\big(\mathbf{H}_{t},\mathbf{H}_{\text{partial}}, \mathbf{M},t\big).
  \label{eq:pred-noise}
\end{equation}
 By minimizing the Kullback-Leibler (KL) divergence between the true posterior $q$ and the learnable distribution $q_{\theta}$, the optimization objective reduces to a mean-matching problem:
\begin{align} 
&\mathop{\arg\!\min}_{\theta}\ \mathrm{KL}\!\big(q\,\|\,q_{\theta}\big)
\notag\\
&= \mathop{\arg\!\min}_{\theta}\ 
\mathbb E_{q}\!\left[\tfrac{1}{2}\big\|\boldsymbol{\mu}(\mathbf H_t,\mathbf H_0,t)
- \boldsymbol{\mu}_{\theta}(\mathbf H_t,\mathbf H_{\text{partial}},t)\big\|_{\boldsymbol{\Sigma}_t^{-1}}^2\right]
\notag\\
&\propto\ \mathop{\arg\!\min}_{\theta}\ 
\mathbb E\!\left[\big\|\boldsymbol{\epsilon}_t
- \mathcal{G}_{\theta}(\mathbf H_t,\mathbf H_{\text{partial}},\mathbf{M}, t)\big\|_2^2\right],
\label{eq:equiv-mean-fitting}
\end{align}
where $\boldsymbol{\epsilon}_t$ is the GT noise derived from the forward process by~\eqref{eq:noisy_channel}.

\paragraph{Non-Markovian Reverse Sampling Process} 
To reduce the latency during the inference phase, we employ a non-Markovian reverse sampling process that significantly reduces the number of required iterations. 
While traditional DDPMs often require hundreds or thousands of sequential denoising steps, leading to prohibitive inference latency, the CDDIM algorithm achieves high-fidelity reconstruction with substantially fewer steps, typically ranging from 20 to 200. 
In this work, we adopt a deterministic non-Markovian reverse sampling trajectory, which is suitable for channel extrapolation since the objective is to accurately reconstruct the missing channel entries conditioned on the observed partial channel.
The reverse sampling trajectory is formulated as
\begin{equation}
\begin{aligned}
\mathbf{H}_{t_1|t_2} 
&= 
\sqrt{\frac{\bar{\alpha}_{t_1}}{\bar{\alpha}_{t_2}}} \mathbf{H}_{t_2} 
+ \Biggl(
\sqrt{1-\bar{\alpha}_{t_1}} 
-
\sqrt{\frac{(1-\bar{\alpha}_{t_2})\bar{\alpha}_{t_1}}{\bar{\alpha}_{t_2}}}
\Biggr) \\
&\quad \times 
\mathcal{G}_{\theta}(\mathbf{H}_{t_2}, \mathbf{H}_{\text{partial}}, \mathbf{M}, t_2),
\end{aligned}
\label{eq:reverse_sampling}
\end{equation}
where $t_1$ and $t_2$ ($t_1 < t_2$) denote two consecutive steps in the accelerated sampling sequence, and $\bar{\alpha}_{t_i} = \prod_{s=1}^{t_i} \alpha_s$ represents the cumulative product of the noise scheduling parameters. 
{\bl The detailed derivation of~\eqref{eq:reverse_sampling} is provided in Appendix~\ref{app:ddim-trajectory}, where the trajectory is obtained by combining the one-step clean-channel estimate with a schedule-consistent non-Markovian noise decomposition.}
By enabling larger reverse-step intervals while preserving the consistency of the marginal distributions, this process accelerates inference by an order of magnitude.

\subsection{Physics-Aware CDDIM Backbone}
\label{subsec:Physics-Aware_Modular_Backbone}

To help exploit the near-field channel correlations while supporting extrapolation across diverse domains, we design a physics-aware Transformer backbone featuring a modular architecture. As illustrated in Fig.~\ref{fig:diffuser}, this framework incorporates universal modular tokenization, condition injection, physics-aware positional encoding, and mask-guided multi-head attention. To enable seamless adaptation to different extrapolation tasks, input-dependent modules are designed to be replaceable (highlighted by red dashed boxes in Fig.~\ref{fig:diffuser}), admitting both 2D and 3D formulations.

\begin{figure*}[t]
    \centering
    \includegraphics[width=0.9\linewidth]{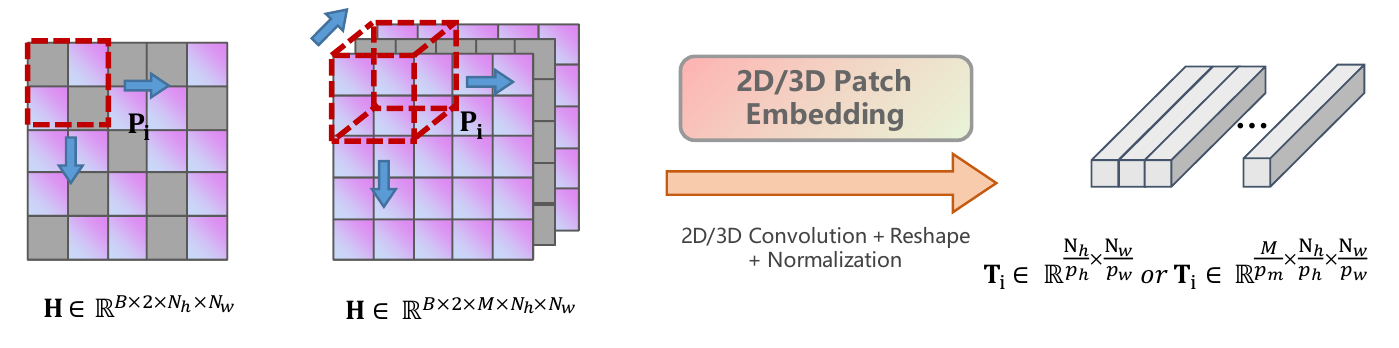}
    \caption{Illustration of the modular input patch-embedding scheme.}
    \label{fig:patch_embedding}
\end{figure*}

\paragraph{Universal Input Representation and Modular Tokenization}
The backbone takes the noisy channel state $\mathbf{H}_t$, the partial observation $\mathbf{H}_{\text{partial}}$, and the sampling mask $\mathbf{M}$ as inputs. The complex-valued channels are decomposed into real and imaginary parts and concatenated into a real-valued tensor.
To achieve domain universality, we employ a modular input patch-embedding module as illustrated in Fig.~\ref{fig:patch_embedding}. While the core Transformer operates on a sequence of flattened tokens $\mathbf{T}_i$, the tokenization interface is tailored to the input dimensionality via a sequence of convolution, reshaping, and normalization as follows:
\begin{itemize}
    \item \textbf{2D Antenna-Domain:} For antenna-domain inputs, the channel tensor $\mathbf{H} \in \mathbb{R}^{B\times 2\times N_h\times N_w}$ is partitioned into patches of size $p_h \times p_w$, where $B$ is the batch size, $p_h$ and $p_w$ are the patch size along the vertical and horizontal axis of the UPA. Each patch $\mathbf{P}_i$ is mapped to a high-dimensional token $\mathbf{T}_i$ via 2D convolution. This results in a flattened token sequence of length $N_{\mathrm{tok}}=\frac{N_h}{p_h}\times\frac{N_w}{p_w}$.
    \item \textbf{3D Frequency-/Spatial-Domains:} For frequency- or spatial-domain inputs, the channel is represented as a volumetric tensor $\mathbf{H} \in \mathbb{R}^{B\times 2\times M \times N_h\times N_w}$. Accordingly, we utilize a 3D convolution with kernel size $p_m \times p_h \times p_w$, where $p_m$ denotes the patch size along the frequency or spatial axis. This yields a token sequence of length $N_{\mathrm{tok}}=\frac{M}{p_m}\times\frac{N_h}{p_h}\times\frac{N_w}{p_w}$.
\end{itemize}

{\bl This design is also consistent with the correlation analysis in Sec.~\ref{sec:Channel Model}-B, where antenna elements, subcarriers, and spatial samples exhibit structured local dependencies. Therefore, the 2D/3D convolutional patch embedding serves as a correlation-aware local tokenization module. Specifically, the 2D patch embedding extracts local antenna-domain features, while the 3D patch embedding extends this tokenization to frequency- or spatial-domain channel tensors by forming frequency-/spatial-indexed antenna patches. In this way, the proposed tokenization preserves the coupling between the two-dimensional antenna-domain structure and the third-dimensional index, rather than treating channel entries as independent samples. The subsequent Transformer layers further learn cross-token and longer-range dependencies along the frequency/spatial and antenna-related dimensions.}
In both cases, the extracted patches are flattened into a token sequence $\mathbf{H}_{\text{embed}} \in \mathbb{R}^{B\times N_{\mathrm{tok}}\times D}$, allowing the subsequent Transformer layers to remain structurally invariant.

\paragraph{Condition Injection}
Task-related side information, including the partial observation $\mathbf{H}_{\text{partial}}$, the mask $\mathbf{M}$, and the diffusion timestep $t$, is injected as conditions to guide the generation. The spatial conditions ($\mathbf{H}_{\text{partial}}, \mathbf{M}$) are processed via a conditional patch-embedding module using convolutional layers with a kernel size equal to the patch size, followed by normalization. Similar to the input module, this extends to 3D convolutions for frequency/spatial tasks to capture inter-domain correlations. The timestep $t$ is separately embedded and added to the tokens, ensuring the reverse sampling process is time-aware.

{\bl 
\paragraph{Positional Encoding}
To provide the Transformer with the position information of channel tokens, we inject sinusoidal positional encodings~\cite{attention} into the token embeddings. This design is motivated by the channel correlation analysis in Sec.~\ref{sec:Channel Model}-B, where the antenna-, frequency-, and spatial-domain correlations are determined by path-length differences and the corresponding phase rotations. In the 2D antenna-domain case, the positional encodings represent the antenna indices on the UPA aperture. Since the array geometry and element spacing are fixed, these indices are directly associated with antenna locations. Under near-field spherical-wave propagation, the UE/scatterer-to-antenna distance varies nonlinearly across the large aperture, and thus the phase and correlation patterns depend on the antenna position. Therefore, PE helps the model distinguish channel patches from different antenna regions and learn their position-dependent correlations. In the 3D frequency- and spatial-domain cases, PE is extended to the joint token positions across the subcarrier/spatial-sampling dimension and the two-dimensional antenna aperture. This helps the model preserve the ordering of frequency/spatial samples and antenna-grid locations, thereby learning the corresponding frequency-antenna and spatial-antenna joint correlations.}

\paragraph{Mask-Guided Multi-Head Attention}
To exploit long-range correlations while adhering to the observations, we introduce a mask-guided attention mechanism. 
Standard self-attention treats all tokens equally, whereas tokens from completely unobserved regions may contain unreliable information and mislead the generation process. 
To mitigate this, we generate a patch-level mask $\mathbf{M}_{\text{P}} \in \mathbb{R}^{B \times N_{\text{tok}}}$ by aggregating the pixel-level mask $\mathbf{M}$. 
Specifically, the average mask value within each patch is first computed. 
A patch is regarded as valid if this average value is larger than zero, indicating that the patch contains at least one observed entry. Otherwise, it is considered fully masked. 
Valid patches are assigned zero attention bias, while fully masked patches are suppressed by a large negative bias $-\infty$ in the attention module.

Given the embedded token sequence $\mathbf{H}_{\text{embed}}$, the query, key, and value tensors are obtained by linear projections
\begin{equation}
\mathbf{Q} = \mathbf{H}_{\text{embed}}\mathbf{W}^Q,\quad
\mathbf{K} = \mathbf{H}_{\text{embed}}\mathbf{W}^K,\quad
\mathbf{V} = \mathbf{H}_{\text{embed}}\mathbf{W}^V,
\end{equation}
where $\mathbf{W}^Q, \mathbf{W}^K, \mathbf{W}^V \in \mathbb{R}^{D \times D}$ denote the learnable weight matrices. Thus, $\mathbf{Q},\mathbf{K},\mathbf{V}\in\mathbb{R}^{B\times N_{\text{tok}}\times D}$. 
For $h$ attention heads, they are reshaped into $\mathbb{R}^{B\times h\times N_{\text{tok}}\times d_k}$, where $d_k=D/h$. 
The patch-level mask $\mathbf{M}_{\text{P}}$ is reshaped to $\mathbb{R}^{B\times1\times1\times N_{\text{tok}}}$ and broadcast to the attention score tensor along the head and query dimensions.
This mask is added to the attention scores before the softmax operation:
\begin{equation}
\text{Attention}(\mathbf{Q}, \mathbf{K}, \mathbf{V}, \mathbf{M}_{\text{P}}) 
= 
\text{softmax}\left(
\frac{\mathbf{Q}\mathbf{K}^{T}}{\sqrt{d_k}} 
+ \mathbf{M}_{\text{P}}
\right)\mathbf{V},
\end{equation}
where the attention score tensor has size $\mathbb{R}^{B\times h\times N_{\text{tok}}\times N_{\text{tok}}}$ after multi-head reshaping. 
The mask is applied along the key-value dimension, so that unknown patches are suppressed when other tokens aggregate contextual information.
This mechanism effectively suppresses attention flow from unobserved regions, forcing the model to attend to reliable observed patches as physical anchors to infer missing components based on learned contextual dependencies.

\begin{algorithm}[t]
\caption{Training Procedure of WGAN-Enhanced CDDIM}
\label{alg:ddim_train}
\begin{algorithmic}[1]
\Require Noisy channel $\mathbf{H}_t$, GT channel $\mathbf{H}_{\mathrm{full}}$, true noise $\boldsymbol{\epsilon}_t$; initialized CDDIM generator $G_\theta$ and Wasserstein critic $D_\phi$; loss weights $\lambda_1,\lambda_2$; clip hyper-parameter $c$; critic steps $n_{\text{critic}}$.
\Ensure Trained generator $\mathcal{G}_{\theta}$ and discriminator $\mathcal{D}_{\phi}$.
\While{$\theta$ has not converged}
  \For{$i=1$ to $n_{\text{critic}}$} \Comment{critic updates}
    \State Predict effective noise via~\eqref{eq:pred-noise} \Comment{$\hat{\boldsymbol{\epsilon}}_t$}
    \State Coarse extrapolation via~\eqref{eq:One-step DDIM} \Comment{$\hat{\mathbf{H}}_{0|t}$}
    \State $\hat{S}_t \leftarrow \mathcal{D}_{\phi}(\hat{\mathbf{H}}_{0|t}),\quad S_t \leftarrow \mathcal{D}_{\phi}(\mathbf{H}_{\text{full}})$
    \State $\mathcal{L}_{\mathcal{D}} \leftarrow \hat{S}_t - S_t$ \Comment{critic loss}
    \State $\phi \leftarrow \phi - \eta_\phi \nabla_\phi \mathcal{L}_{\mathcal{D}}$ \Comment{critic weight updates}
    \State \textbf{clip} $\phi$ element-wise to $[-c,c]$ \Comment{weight clipping}
  \EndFor
  \State \Comment{generator (diffuser) update}
  \State Predict noise and coarse extrapolation again to form $\hat{\mathbf{H}}_{0|t}$ via~\eqref{eq:pred-noise} and~\eqref{eq:One-step DDIM}
  \State $\hat{S}_t \leftarrow \mathcal{D}_{\phi}(\hat{\mathbf{H}}_{0|t})$
  \State $\mathcal{L}_{\mathcal{G}} \leftarrow \lambda_1 \cdot \mathcal{L}_n + \lambda_2 \cdot \mathcal{L}_h - \hat{S}_t$ \Comment{generator loss in~\eqref{eq:loss}}
  \State $\theta \leftarrow \theta - \eta_\theta \nabla_\theta \mathcal{L}_{\mathcal{G}}$ \Comment{generator weight updates}
\EndWhile
\end{algorithmic}
\end{algorithm}

\subsection{WGAN-Enhanced CDDIM Channel Extrapolator}
The architecture of the WGAN-enhanced CDDIM channel extrapolator is illustrated in Fig.~\ref{fig:diffuser}, which comprises the proposed physics-aware CDDIM generator explained in Sec.~\ref{subsec:Physics-Aware_Modular_Backbone} and an accompanying WGAN-based discriminator.
\paragraph{WGAN Integration and Adversarial Supervision} To further improve the quality of the generated channels and provide adversarial feedback to the diffusion-based generator, WGAN discriminator is introduced. The training procedure of the WGAN-enhanced CDDIM framework is summarized in Algorithm~\ref{alg:ddim_train}. During each training iteration, given a noisy observation $\mathbf{H}_t$ and its corresponding predicted noise $\hat{\boldsymbol{\epsilon}}_t$, the generator performs a one-step DDIM reverse approximation to derive a coarse estimation of the full channel by:

\begin{equation}
    \hat{\mathbf{H}}_{0|\text{t}} = \frac{\mathbf{H}_t - \sqrt{1 - \bar{\alpha}_t} \cdot \hat{\boldsymbol{\epsilon}}_t}{\sqrt{\bar{\alpha}_t}}.
\label{eq:One-step DDIM}
\end{equation}
The discriminator $\mathcal{D}_{\phi}$ is introduced as a WGAN critic to measure the distributional discrepancy between the GT channel and the coarsely reconstructed channel. 
Given the critic scores
\begin{equation}
S_t=\mathcal{D}_{\phi}(\mathbf{H}_{\mathrm{full}}), 
\qquad
\hat{S}_t=\mathcal{D}_{\phi}(\hat{\mathbf{H}}_{0|t}),
\end{equation}
where $\mathbf{H}_{\mathrm{full}}$ and $\hat{\mathbf{H}}_{0|t}$ denote the GT channel and the one-step reconstructed channel at reverse step $t$, respectively, the critic is trained to enlarge their Wasserstein distance. 
Equivalently, when using a minimization objective, the discriminator loss is
\begin{equation}
\mathcal{L}_{\mathcal{D}}=\hat{S}_t-S_t .
\end{equation}
Following~\cite{arjovsky2017wgan}, the $K$-Lipschitz constraint is approximately enforced by clipping the parameters of $\mathcal{D}_{\phi}$ to a compact range $[-c,c]$ after each discriminator update. 
Moreover, the discriminator is updated $n_{\mathrm{critic}}$ times for each generator update, so that the critic can provide a more reliable Wasserstein gradient for guiding the CDDIM-based generation process.

The generator is optimized by a weighted-sum loss function
\begin{equation}
    \mathcal{L}_{\mathcal{G}} = \lambda_1 \cdot \mathcal{L}_n + \lambda_2 \cdot \mathcal{L}_h - \hat{S}_t,
\label{eq:loss}
\end{equation}
where $\mathcal{L}_n= \| \hat{\boldsymbol{\epsilon}}_t - \boldsymbol{\epsilon}_t \|_2^2$ constrains the noise prediction, and $\mathcal{L}_h = \| \hat{\mathbf{H}}_{0|t} - \mathbf{H}_{\text{full}} \|_2^2$ constrains the coarse channel reconstruction. The term $-\hat{S}_t$ acts as an adversarial incentive derived from the WGAN critic, encouraging the generator to improve distributional fidelity by increasing the critic score.

\begin{algorithm}[t]
\caption{Reverse Sampling with Discriminator Guidance}
\label{alg:ddim-discriminator}
\begin{algorithmic}[1]
\Require Initial noise $\mathbf H_{T-1}$, observed partial channel $\mathbf H_{\rm partial}$, mask $\mathbf M$, mask ratio $\gamma$, pre-trained DDIM network $G_\theta$, discriminator $D_\phi$, noise schedule $\{\bar\alpha_t\}$, sampling interval $\Delta t$, guidance step size $\delta$, and RePaint threshold $\gamma_{\rm thre}$.
\Ensure Extrapolated channel $\hat{\mathbf{H}}_0$.
\For{$t = T-1;\ t \ge 0;\ t \leftarrow t - \Delta t$}
    \State Predict the effective noise at step $t$ via~\eqref{eq:pred-noise} \Comment{$\hat{\boldsymbol{\epsilon}}_{t}$}
    \State Coarsely extrapolate the full channel using the current predicted channel $\hat{\mathbf{H}}_{t}$ via~\eqref{eq:One-step DDIM} \Comment{$\hat{\mathbf{H}}_{0|t}$}
    \State Compute discriminator gradient and update the effective noise via~\eqref{eq:updated_noise} \Comment{refined $\hat{\boldsymbol{\epsilon}}_{t}$}
    \State Reverse sampling to the previous step via~\eqref{eq:reverse_sampling} \Comment{$\hat{\mathbf{H}}_{t-\Delta t}$}
    \If{$\gamma < \gamma_{\text{thre}}$} \Comment{{\it $\gamma$-gated RePaint}}
        \State Apply clean RePaint via~\eqref{eq:cleanoverwrite}
    \Else
        \State Apply noisy RePaint via~\eqref{eq:noisy_known_channel} and~\eqref{eq:noisy_repaint}
    \EndIf
\EndFor
\State Apply the last clean RePaint to $\hat{\mathbf{H}}_0$ via~\eqref{eq:cleanoverwrite}.
\Comment{ $\hat{\mathbf{H}}_0$}
\end{algorithmic}
\end{algorithm}

\paragraph{Discriminator-Guided Reverse Sampling}
To further enhance the generation quality during the inference phase, we incorporate a discriminator-guided correction mechanism into the reverse sampling trajectory, as illustrated in Algorithm~\ref{alg:ddim-discriminator}. Drawing inspiration from the score distillation sampling~\cite{poole2023dreamfusion}, the pre-trained discriminator is utilized as a learned proxy of the underlying channel distribution. By leveraging the gradient field of this proxy, the model can iteratively refine the denoising path to better align with the characteristics of realistic channels. Since the discriminator $\mathcal{D}_{\phi}$ is trained to assign higher scores to authentic channel samples, its gradient serves as an implicit quality indicator, providing a ``plug-and-play" guidance signal during sampling.
{\bl Specifically, after predicting the effective noise $\hat{\boldsymbol{\epsilon}}_{t}$ 
at timestep $t$ by~\eqref{eq:pred-noise}, we first obtain the coarsely 
extrapolated channel $\hat{\mathbf{H}}_{0|t}$ via~\eqref{eq:One-step DDIM} and 
the corresponding discriminator score 
$\hat{S}_{t} = \mathcal{D}_{\phi}(\hat{\mathbf{H}}_{0|t})$. 
Since $\mathbf{H}_t$ is fixed when applying the discriminator guidance at step $t$, 
the Jacobian of $\hat{\mathbf{H}}_{0|t}$ with respect to 
$\hat{\boldsymbol{\epsilon}}_t$ is given by
\begin{equation}
\frac{\partial \hat{\mathbf{H}}_{0|t}}
{\partial \hat{\boldsymbol{\epsilon}}_t}
=
-
\sqrt{\frac{1-\bar{\alpha}_t}{\bar{\alpha}_t}}\mathbf{I}.
\end{equation}
Therefore, by the chain rule, the gradient of the discriminator score with 
respect to the predicted noise is
\begin{equation}
\nabla_{\hat{\boldsymbol{\epsilon}}_t}\hat{S}_t
=
-
\sqrt{\frac{1-\bar{\alpha}_t}{\bar{\alpha}_t}}
\nabla_{\hat{\mathbf{H}}_{0|t}}
\mathcal{D}_{\phi}(\hat{\mathbf{H}}_{0|t}).
\end{equation}
Accordingly, to increase the discriminator score and steer the generated 
channel toward a more realistic region, the predicted noise is refined by
\begin{align}
\hat{\boldsymbol{\epsilon}}_{t}
&\gets
\hat{\boldsymbol{\epsilon}}_{t}
+
\delta\,\nabla_{\hat{\boldsymbol{\epsilon}}_{t}}\hat{S}_t \nonumber\\
&=
\hat{\boldsymbol{\epsilon}}_{t}
-
\delta
\sqrt{\frac{1-\bar{\alpha}_t}{\bar{\alpha}_t}}
\nabla_{\hat{\mathbf{H}}_{0|t}}
\mathcal{D}_{\phi}(\hat{\mathbf{H}}_{0|t}),
\label{eq:updated_noise}
\end{align}
where $\delta>0$ is the guidance step size.}

For the DDIM reverse sampling regime, we define an accelerated sampling strategy utilizing a subset of diffusion steps $\{t_0, t_1, \ldots, t_K\}$, where the number of steps is significantly reduced ($K \ll T$), and $t_0 = 0$, $t_K=T-1$. By setting the step interval to $\Delta_t$ and performing non-Markovian reverse sampling only at these discrete intervals, the reverse sampling is accelerated by an order of magnitude. The intermediate channel state $\hat{\mathbf{H}}_{t-\Delta t}$ is iteratively inferred from $\mathbf{H}_{t}$ according to the deterministic trajectory defined in \eqref{eq:reverse_sampling}.
To balance the precision of physical measurements with the numerical stability of the generative process, we employ a RePaint-style~\cite{lugmayr2022repaint} conditioning strategy, refined into a $\gamma$-gated RePaint during the reverse sampling. The key idea is to let the channel overwrite rule depend on the mask ratio threshold~$\gamma_{\text{thre}}$.

In the \textbf{Low-mask RePaint regime (Clean RePaint)}, where $\gamma < \gamma_{\text{thre}}$, the observations already provide a strong and reliable constraint for extrapolation. In this regime, we directly overwrite the known locations of channels with the clean partial channel at each reverse step to eliminate noise-induced residuals by
\begin{equation}
    \hat{\mathbf{H}}_{t-\Delta t}
    \leftarrow
    \mathbf{M}\odot \mathbf{H}_{\text{partial}}
    \;+\;
    (\mathbf{1}-\mathbf{M})\odot \hat{\mathbf{H}}_{t-\Delta t}.
\label{eq:cleanoverwrite}
\end{equation}
Conversely, in the \textbf{High-mask RePaint regime (Noisy RePaint)} occurring when $\gamma \ge \gamma_{\text{thre}}$, the sparse observations are insufficient to guide the trajectory alone, and direct insertion of clean values would induce a distributional mismatch with the noisy generative manifold at step $t$. To mitigate this, following RePaint~\cite{lugmayr2022repaint}, we re-noise the observed partial channel to the current diffusion level before overwriting. Specifically, the known components are first mapped to the noise level of timestep $t-\Delta t$ via~\eqref{eq:noisy_channel} to obtain
\begin{equation}
\mathbf H_{t-\Delta t,\rm partial}
=
\sqrt{\bar\alpha_{t-\Delta t}}\mathbf H_{\rm partial}
+
\sqrt{1-\bar\alpha_{t-\Delta t}}\mathbf z_{t-\Delta t},
\label{eq:noisy_known_channel}
\end{equation}
where $\mathbf z_{t-\Delta t} \sim \mathcal{CN}(\mathbf 0,\mathbf I)$.
The estimated channel is updated by
\begin{equation}
    \hat{\mathbf{H}}_{t-\Delta t}
    \leftarrow
    \mathbf{M}\odot \mathbf{H}_{t-\Delta t,\text{partial}}
    \;+\;
    (\mathbf{1}-\mathbf{M})\odot \hat{\mathbf{H}}_{t-\Delta t}.
\label{eq:noisy_repaint}
\end{equation}
This schedule-aligned overwrite keeps the known regions at the target reverse-step noise level, prevents distributional drift, and helps propagate information from scarce observations in a statistically consistent way.
In the final inference step, we apply the clean RePaint via~\eqref{eq:cleanoverwrite} to finally overwrite the known parts with the observed channel and obtain the extrapolated channel $\hat{\mathbf{H}}_0$.

\section{Experimental Results}
\label{sec:experiments}

In this section, we evaluate the proposed WGAN-enhanced CDDIM framework for near-field XL-MIMO channel extrapolation. 
We first describe the simulation setup, evaluation metrics, and baseline methods. 
Then, we empirically examine the channel correlations in the antenna, frequency, and spatial domains to validate the multi-domain dependencies discussed in Sec.~\ref{sec:Channel Model}. 
Based on these observations, we conduct detailed performance comparisons for antenna-domain extrapolation. 
Finally, we investigate the generalization capability of the proposed framework and further extend the evaluation to frequency- and spatial-domain extrapolation.

\subsection{Experimental Setups}

The simulation scenarios and hyperparameter configurations are summarized in Table~\ref{tab:config}. 
Unless otherwise specified, the channel datasets are simulated using QuaDRiGa
under the 3GPP urban microcell (UMi) scenario~\cite{jaeckel2014quadriga,3gpp38901}. In the simulation,
the BS UPA is modeled with explicit element positions, and the channel
coefficient of each antenna element is generated according to its own geometric
relationship with the LoS component and the multipath clusters, which is consistent with the spherical-wave
representation in Sec.~\ref{sec:Channel Model}.

The default training data are generated under the near-field setting with a BS equipped with a $32\times32$ UPA operating at $f_c=7\,\mathrm{GHz}$, where single-antenna users are randomly distributed within the Rayleigh distance. To examine the robustness of the proposed framework, we further evaluate it under different carrier frequencies, array sizes, and near-/far-field regimes.
For each scenario, the dataset contains 10000 training samples, 1000 validation samples, and 200 testing samples. All channel samples are standardized using Z-score normalization based on the training-set mean and standard deviation, denoted by $\mu_{\text{train}}$ and $\sigma_{\text{train}}$, respectively, i.e.,
$\tilde{\mathbf H}=(\mathbf H-\mu_{\text{train}})/\sigma_{\text{train}}$. The proposed backbone is implemented with a patch-based Transformer backbone, where the domain-specific patch sizes, hidden dimensions, and loss weights are listed in Table~\ref{tab:config}. The training and diffusion-related parameters are also reported in Table~\ref{tab:config}. All experiments are conducted on an NVIDIA GeForce RTX 4090 GPU.

\begin{table}[t]
    \centering
    \footnotesize
    \caption{Experimental setups.}
    \label{tab:config}
    \begin{tabular}{p{0.42\linewidth} p{0.52\linewidth}}
        \toprule
        Category & Configuration \\
        \midrule
        \multicolumn{2}{l}{\textbf{Scenario and dataset}} \\
scenario & 3GPP UMi\\
        Training central frequency & $f_c = 7\,\mathrm{GHz}$ \\
        Test central frequencies & $f_c \in \{6,7,8\}\,\mathrm{GHz}$ \\
        Bandwidth & $200\,\mathrm{MHz}$ \\
        Subcarriers & $M=32$ ($\Delta f = 6.25\,\mathrm{MHz}$) \\
        BS array (near field) & UPA $32\times32$ \\
        Spatial samples & $U = 32$ \\
        Spatial sampling interval & 0.1\,m\\
        Rayleigh distance & $D_R = 43.9$\,m \\
        Train / validation / test samples & 10000 / 1000 / 200 \\
        Additional arrays (far field) & UPA $16\times16$, $32\times32$ \\

        \addlinespace[0.38em]
        \multicolumn{2}{l}{\textbf{Training and diffusion}} \\
        Epochs / batch size & 2000 / 256 \\
        Optimizer & AdamW, initial learning rate $2\times10^{-4}$ \\
        Diffusion steps  & $T=1000$ \\
        Noise schedule & Linear $\beta_t \in [10^{-4},\,2\times10^{-2}]$ \\
        DDIM sampling steps  & $K=100$ \\
        Patch size (antenna) & $p_{\mathrm{h}}=8$, $p_{\mathrm{w}}=8$ \\
        Patch size (frequency/spatial) & $p_{\mathrm{m}}=1$, $p_{\mathrm{h}}=8$, $p_{\mathrm{w}}=8$ \\
        Number of attention heads  & $h=16$ \\
        Hidden dimension  & $D=256$ (antenna), 384 (frequency/spatial) \\
        Loss weights $(\lambda_1,\lambda_2)$ & (500, 100) antenna; (20, 20) frequency/spatial \\
        WGAN clipping  & $c=0.01$ \\
        Critic iterations  & $n_{\mathrm{critic}}=5$ \\
        Guidance strength  & $\delta=0.1$ \\
        Mask ratio & $\gamma$ is sampled from $\{0.2,0.3,\ldots,0.8\}$ \\
        RePaint threshold  & $\gamma_{\mathrm{thre}}=0.7$ (random mask), $\gamma_{\mathrm{thre}}=0.5$ (structured mask) \\
        \bottomrule
    \end{tabular}
\end{table}

To evaluate the extrapolation performance, we adopt three evaluation metrics, including normalized mean squared error (NMSE), $L_1$ distance, and cosine distance. 
{\bl NMSE reflects the normalized reconstruction error, $L_1$ distance characterizes the element-wise absolute deviation, and cosine distance evaluates the global structural similarity between the extrapolated and GT channels. 
These metrics are jointly used to assess extrapolation performance from complementary perspectives, including energy-level accuracy, local reconstruction error, and structural consistency.}
In the angular domain, we additionally evaluate the mean absolute error (MAE) and mean squared error (MSE) for pixel-wise errors, the peak signal-to-noise ratio (PSNR) for peak-normalized reconstruction quality, and the correlation coefficient to quantify structural similarity. 
The following representative baselines are included for comparison:

\begin{itemize}
    \item \textbf{Gaussian Filling (Random):} 
    Masked entries are filled with i.i.d. complex Gaussian noise by $\mathcal{CN}(\mu_{\text{train}},\,\sigma_{\text{train}}^{2})$.

    \item \textbf{Partial Channel with Zero Filling (Partial):} 
    A naive baseline that directly fills the masked locations with zeros, serving as a lower bound for performance.

    \item \textbf{Inverse Distance Weighting (IDW) Interpolation:} 
    A classical deterministic algorithm that estimates missing values based on the weighted average of observed neighbors. It serves as a heuristic baseline specifically for structured frequency and spatial masks.

    {\bl \item \textbf{Orthogonal Matching Pursuit (OMP)~\cite{tropp2007omp}:}
    A conventional compressed sensing baseline for antenna-domain extrapolation. 
    It reconstructs the full channel from partial antenna observations via iterative codeword search and least-squares coefficient update. 
     } 

    \item \textbf{Conditional WGAN (WGAN)~\cite{balevi2021wcgan}:} 
    A Wasserstein GAN-based extrapolator that utilizes the Wasserstein distance as the objective function to enhance training stability and generative quality compared to standard GANs.

    \item \textbf{Conditional DDPM (CDDPM)~\cite{zhang2025cddpm}:} 
    A conditional diffusion-based benchmark that employs a Transformer backbone for noise prediction and performs channel reconstruction through the standard Markovian reverse denoising process. It serves as a representative conventional diffusion model, without the physics-aware design and non-Markovian sampling acceleration introduced in the proposed framework.

    \item \textbf{Physics-Aware CDDIM (Ablation of WGAN):} 
    An ablation variant of the proposed framework that preserves the physics-aware backbone and non-Markovian CDDIM sampling, while excluding the WGAN-guided adversarial training and reverse refinement mechanism. This baseline helps quantify the contribution of the adversarial guidance module.
\end{itemize}

{\bl 
\subsection{Channel Correlation Analysis}

For empirical validation, the correlation in \eqref{eq:corr_def} is approximated by finite-sample averaging on the generated test dataset. 
The antenna-domain correlation is computed over antenna-index differences at the center subcarrier, while the frequency- and spatial-domain correlations are evaluated using full-array channel responses over subcarrier and spatial-point pairs, respectively. 
The joint correlations are obtained by additionally introducing antenna-index differences into the frequency- and spatial-domain correlation calculations. 
For the spatial-domain and spatial-antenna joint results, the correlations are averaged within distance bins.

\begin{figure*}[t]
\centering
\includegraphics[width=0.99\linewidth]{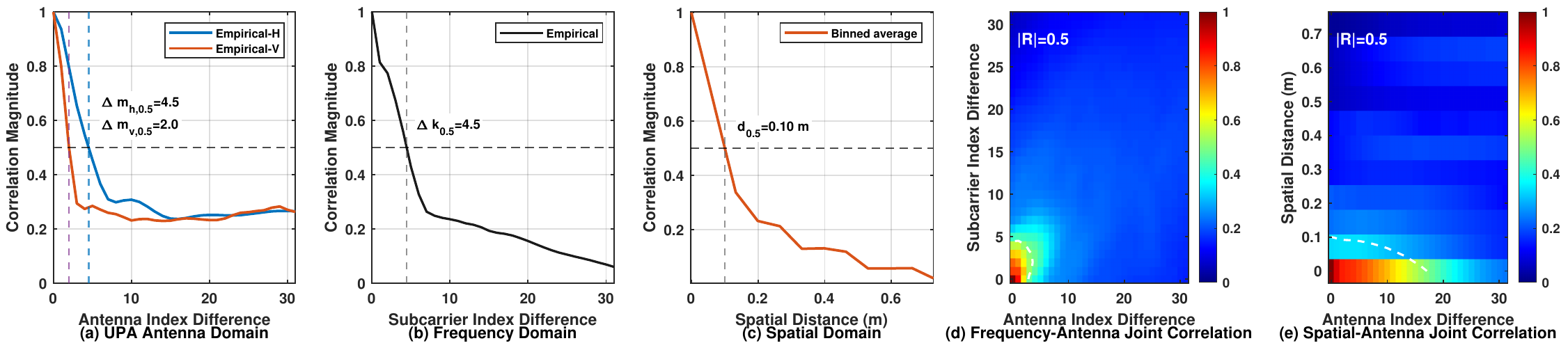}
\caption{{\bl Empirical channel correlations in the (a) antenna, (b) frequency, (c) spatial, and (d,e) joint domains. The dashed lines and white contours indicate the half-correlation boundary $|\mathrm{R}|=0.5$.}}
\label{fig:multi_domain_correlation}
\end{figure*}

\begin{figure*}[t]
\centering

\newcommand{\w}{0.333\linewidth}   
\newcommand{\colgap}{-0.2em}       
\newcommand{\rowgap}{-0.3em}        

\subfigure[NMSE under random mask]{%
    \includegraphics[width=\w]{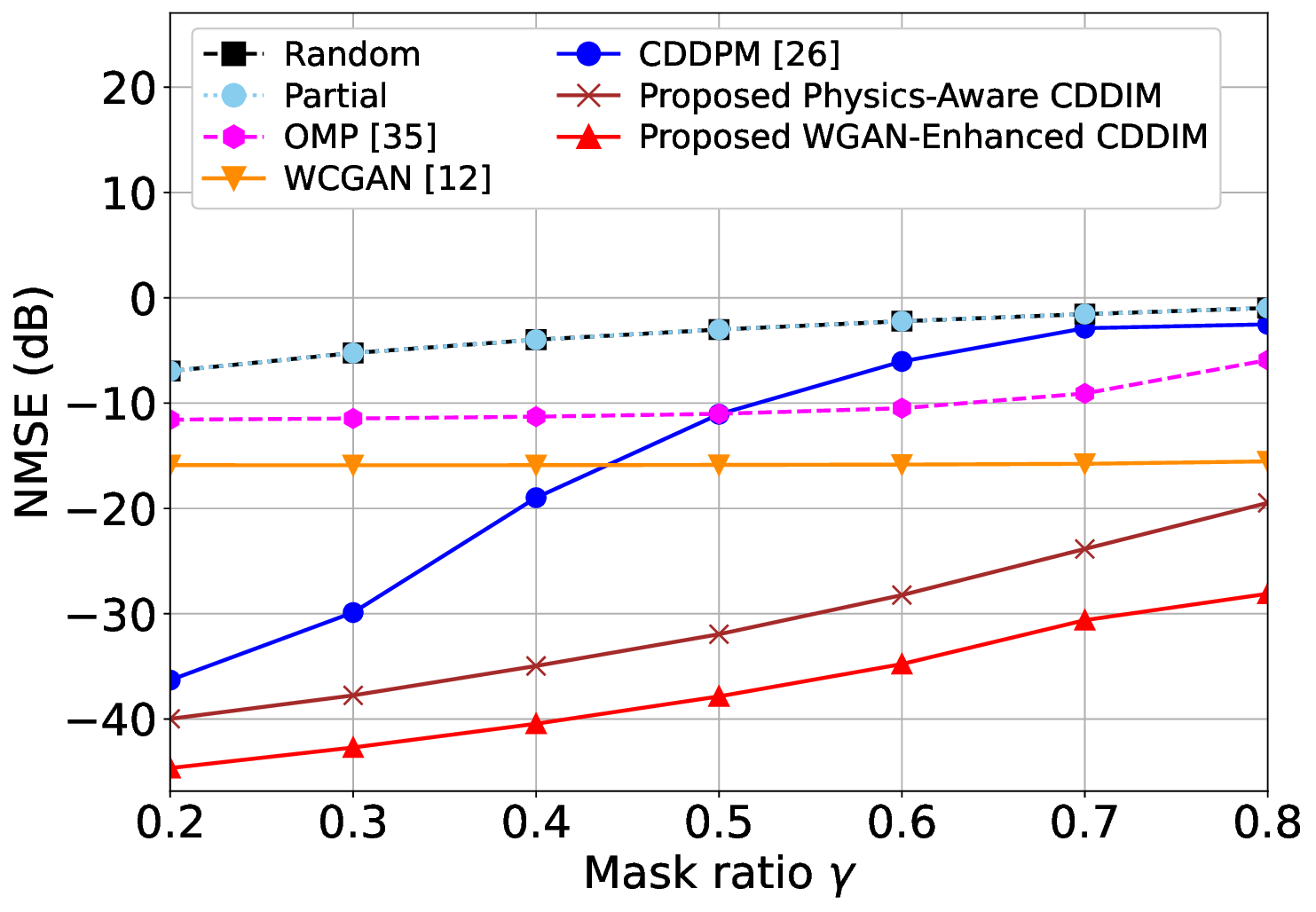}%
}\hspace{\colgap}%
\subfigure[$L_1$ distance under random mask]{%
    \includegraphics[width=\w]{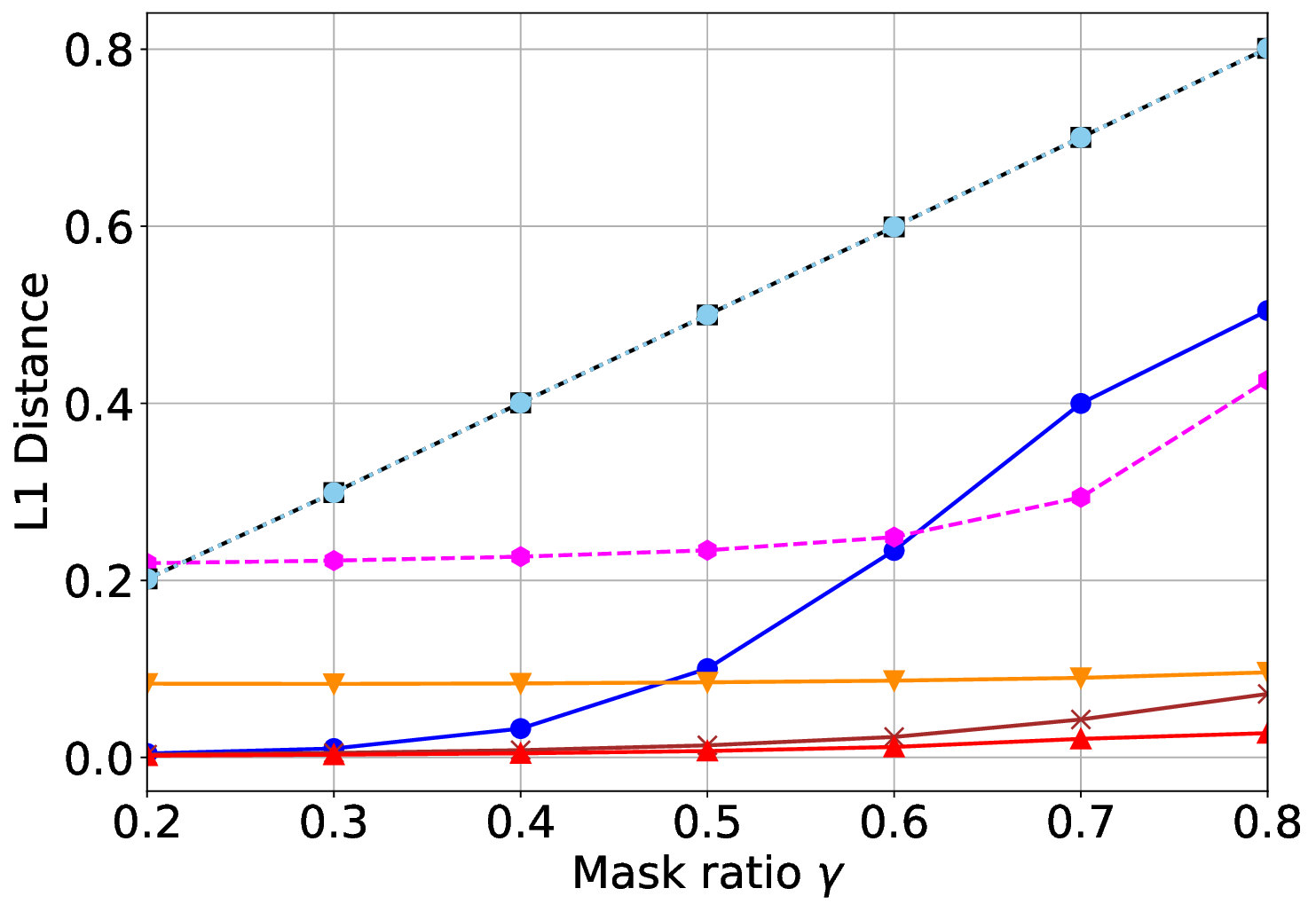}%
}\hspace{\colgap}%
\subfigure[Cosine distance under random mask]{%
    \includegraphics[width=\w]{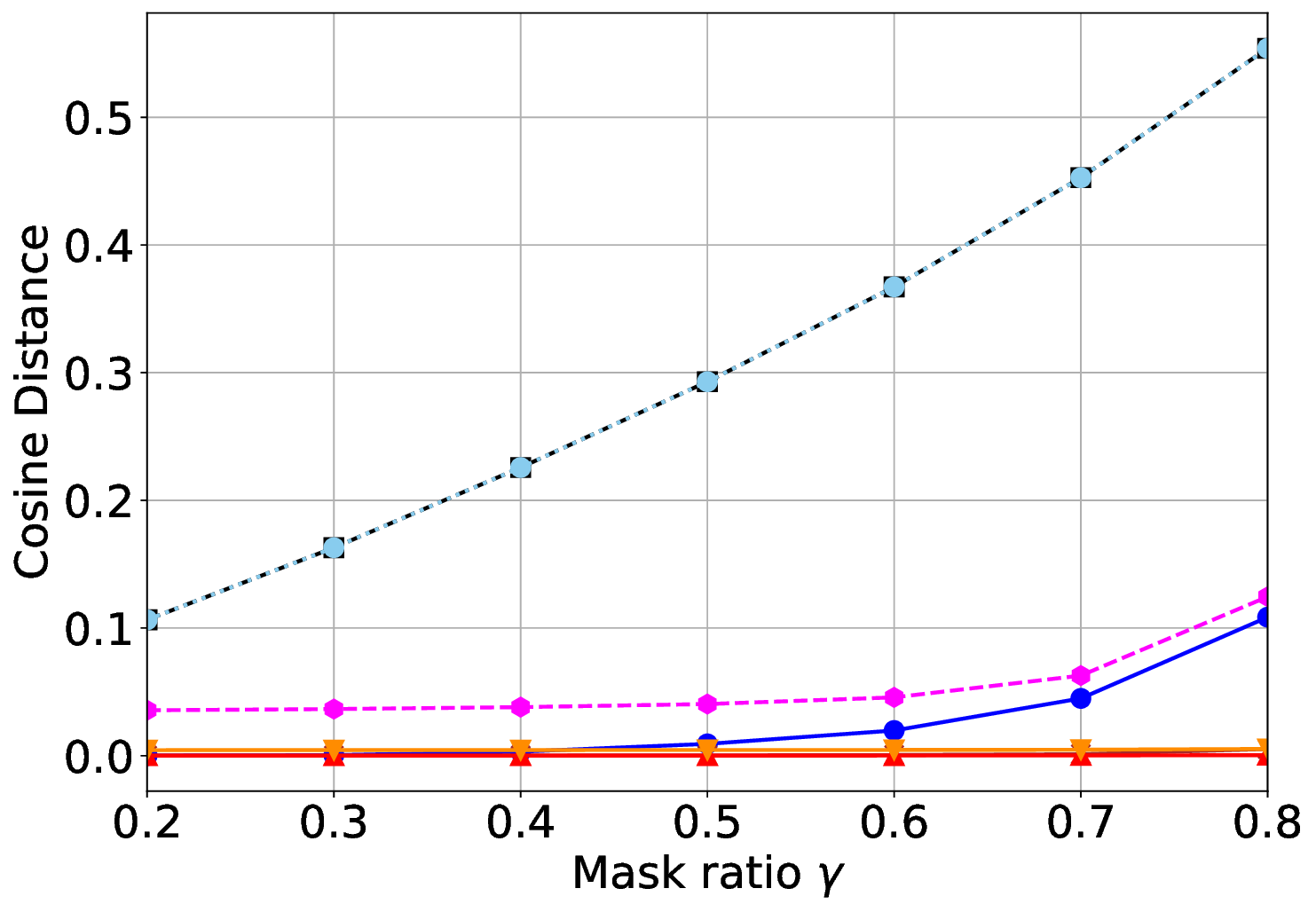}%
}

\vspace{\rowgap} 

\subfigure[NMSE under structured mask]{%
    \includegraphics[width=\w]{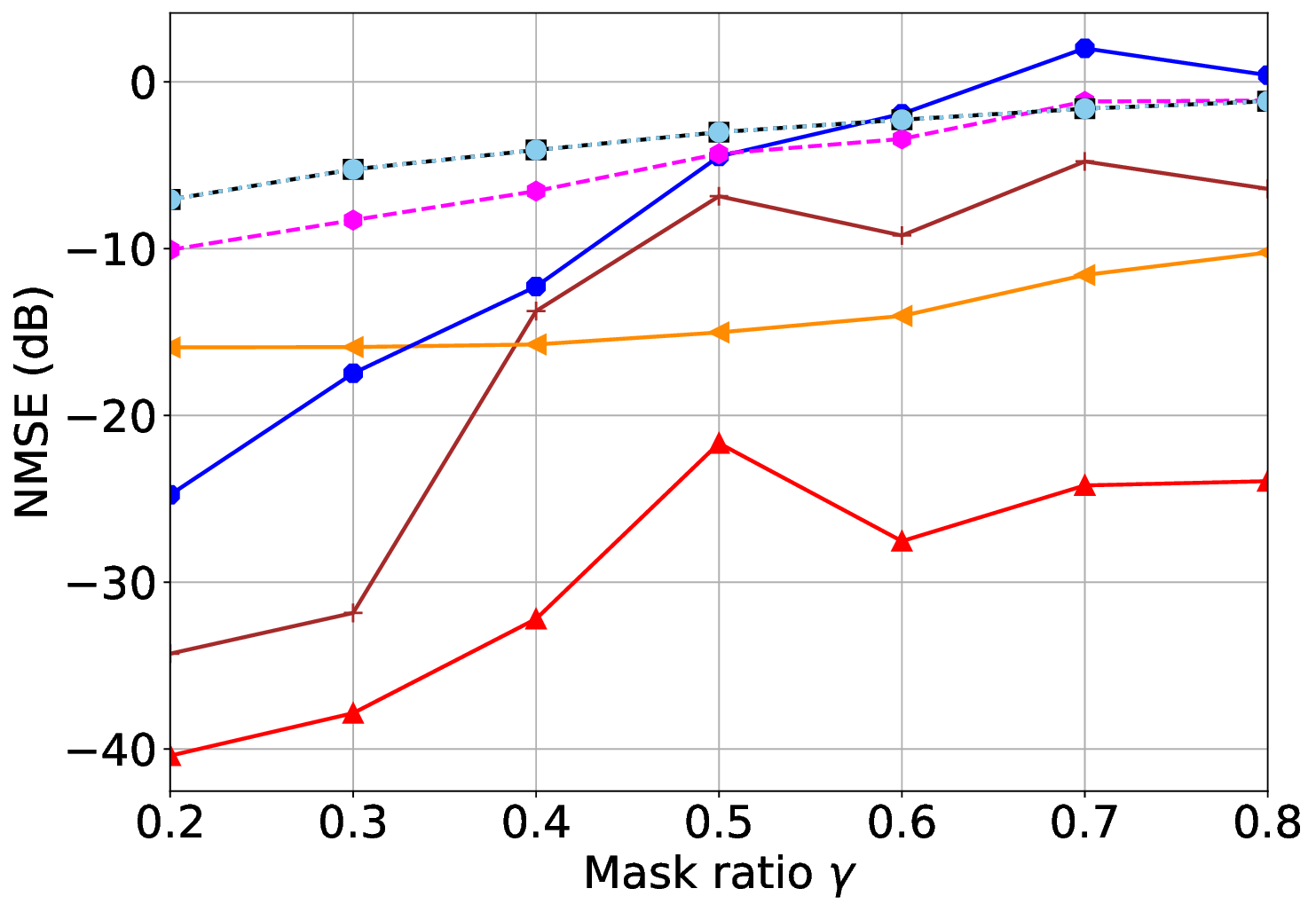}%
}\hspace{\colgap}%
\subfigure[$L_1$ distance under structured mask]{%
    \includegraphics[width=\w]{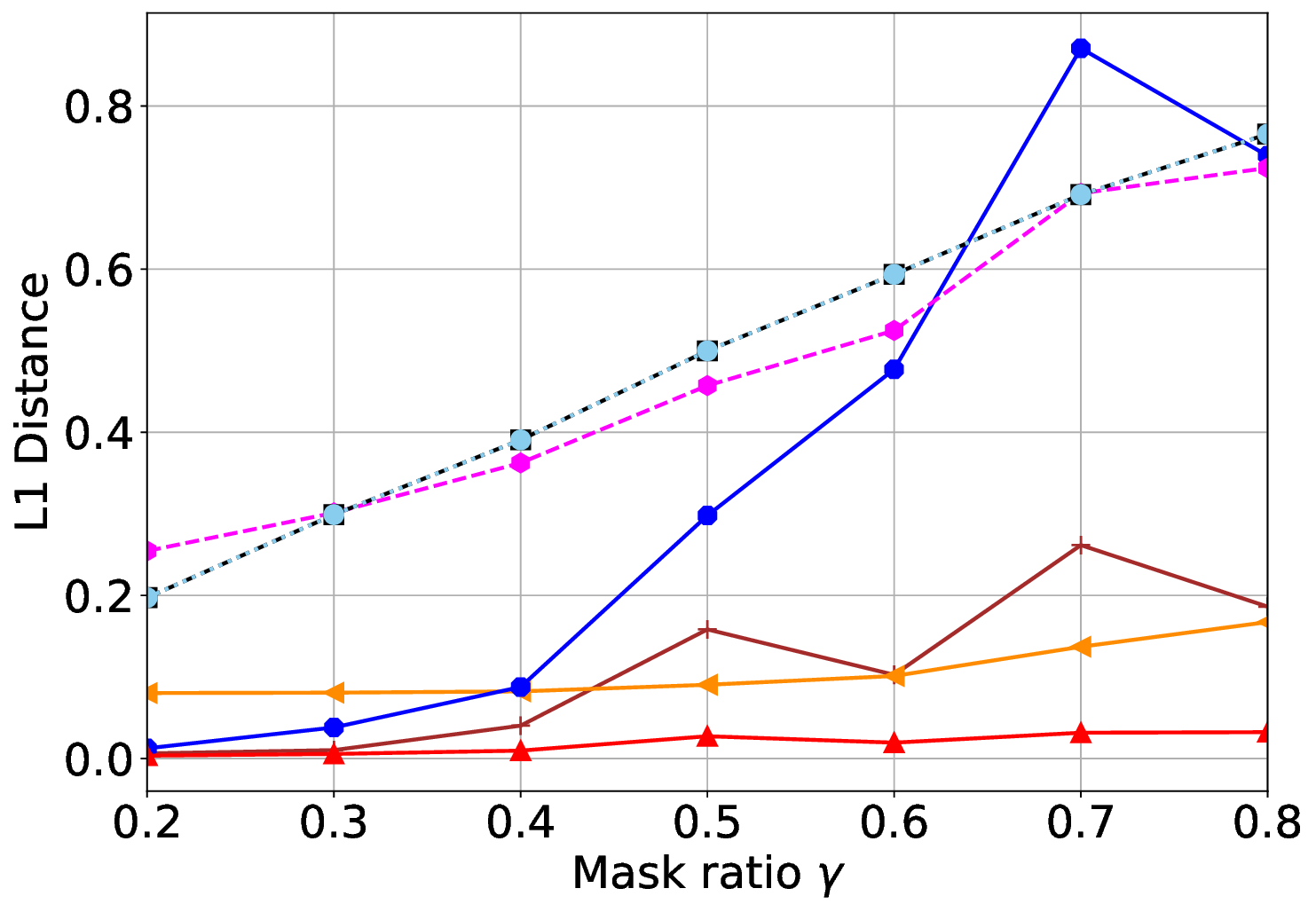}%
}\hspace{\colgap}%
\subfigure[Cosine distance under structured mask]{%
    \includegraphics[width=\w]{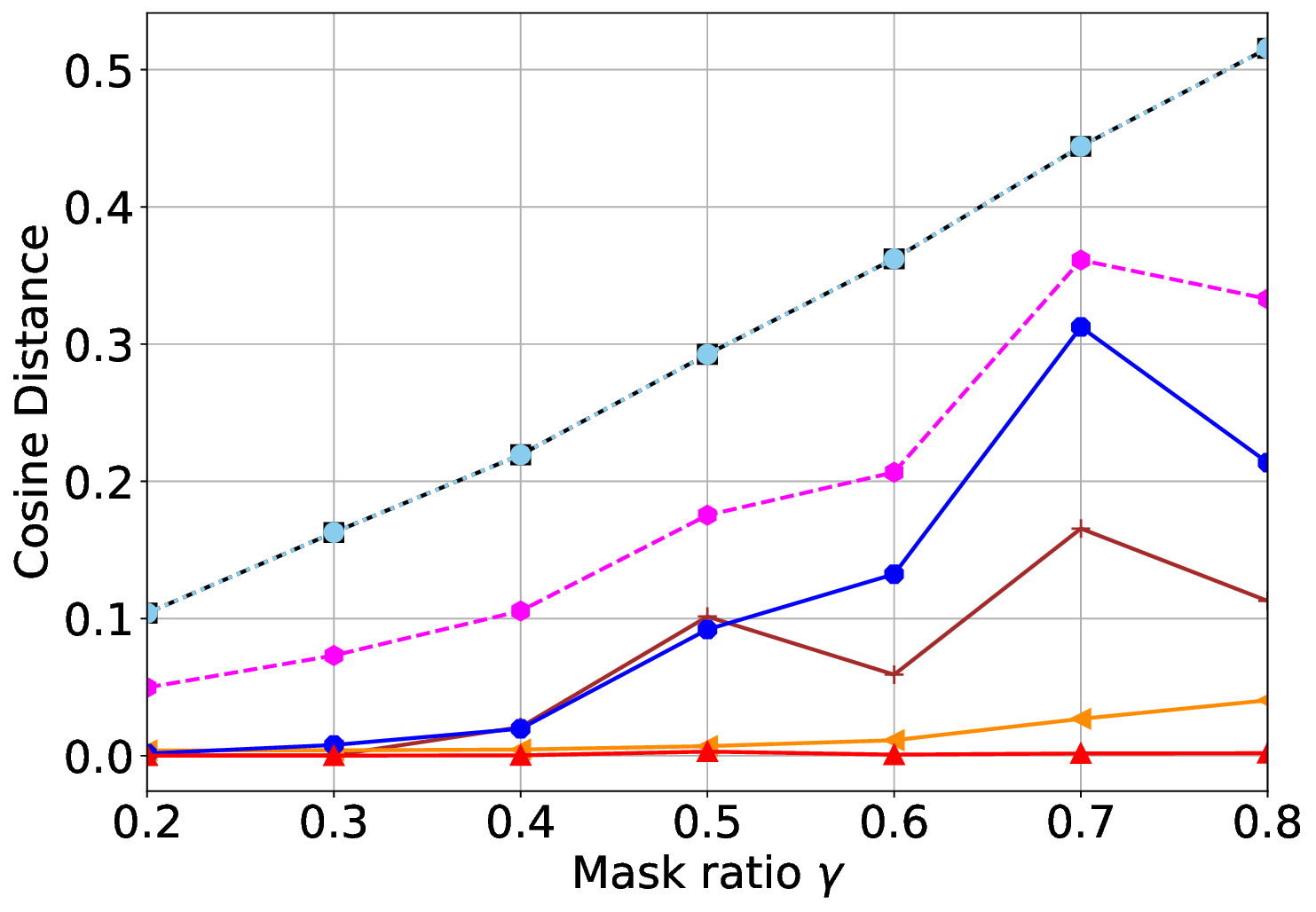}%
}

\caption{{\bl Channel extrapolation performance comparison under antenna-domain masking. Top row: performance under random mask: (a) NMSE, (b) $L_1$ distance, (c) cosine distance. Bottom row: performance under block-structured masking: (d) NMSE, (e) $L_1$ distance, (f) cosine distance.}}
\label{fig:antenna_results}
\end{figure*}

As shown in Fig.~\ref{fig:multi_domain_correlation}(a-c), the correlations in different domains generally decrease as the corresponding spacing increases. 
Specifically, in the antenna domain, the correlation decreases with the antenna-index spacing due to aperture-dependent phase variations caused by spherical wavefront propagation. 
In the frequency domain, the correlation gradually decreases with the subcarrier spacing because of multi-path delay differences. 
In the spatial domain, the correlation also decreases with the physical distance between spatial samples, suggesting that nearby samples are more informative for spatial-domain extrapolation. The marked half-correlation extents quantify these local dependency ranges, with empirical values of $\Delta m_{\mathrm{h},0.5}\approx 4.5$, $\Delta m_{\mathrm{v},0.5}\approx 2.0$, $\Delta k_{0.5}\approx 4.5 (28.125$ MHz), and $d_{0.5}\approx 0.1$ m.
Figs.~\ref{fig:multi_domain_correlation}(d) and (e) further show that strong joint correlations are mainly concentrated around small frequency/antenna or spatial/antenna separations. These results verify the existence of local channel dependencies across neighboring antennas, subcarriers, and spatial samples. Therefore, the proposed patch embedding is physically motivated, as it encodes locally correlated channel entries into tokens, while the subsequent Transformer blocks further learn cross-token and longer-range dependencies for multi-domain channel extrapolation.}

\begin{figure*}[t]
\centering

\newcommand{\w}{0.333\linewidth}
\newcommand{\colgap}{-0.2em}

\subfigure[$\mathrm{SNR}=-10$ dB]{%
    \includegraphics[width=\w]{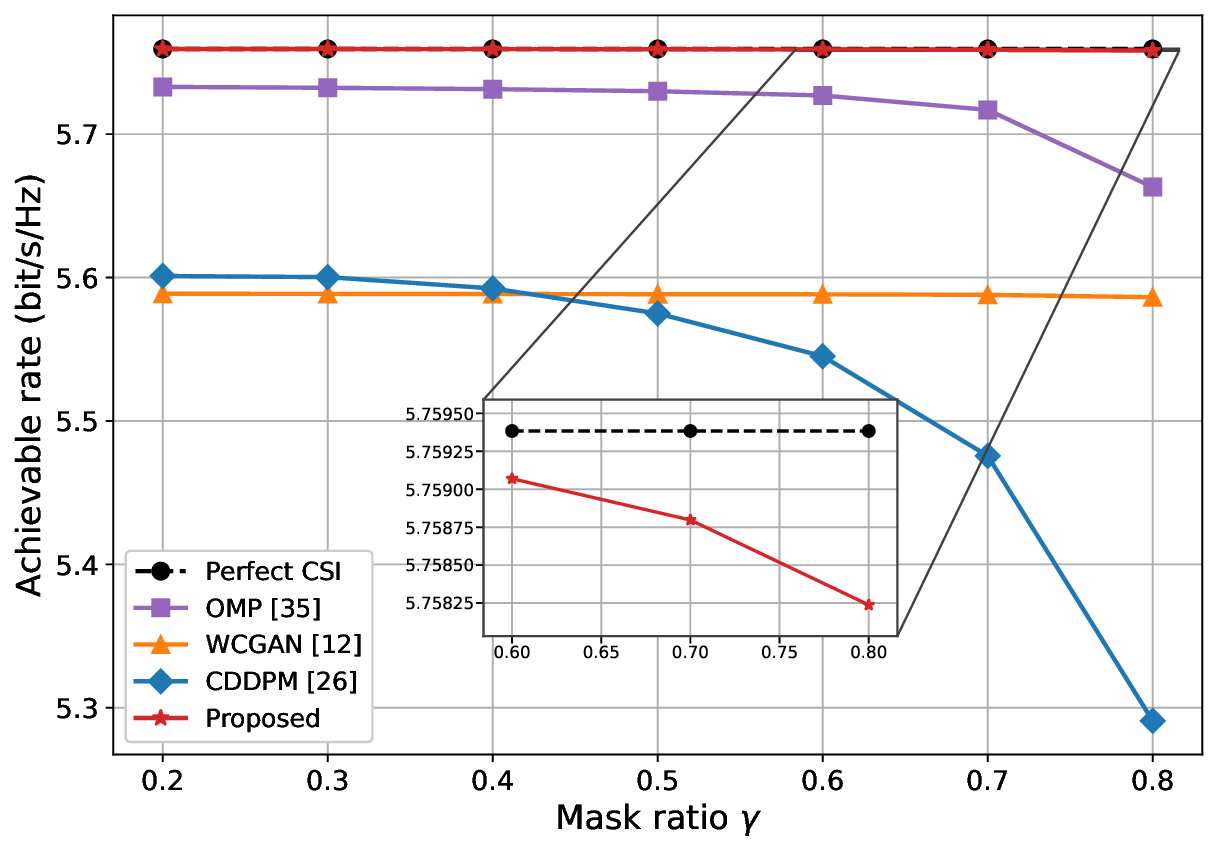}%
}\hspace{\colgap}%
\subfigure[$\mathrm{SNR}=0$ dB]{%
    \includegraphics[width=\w]{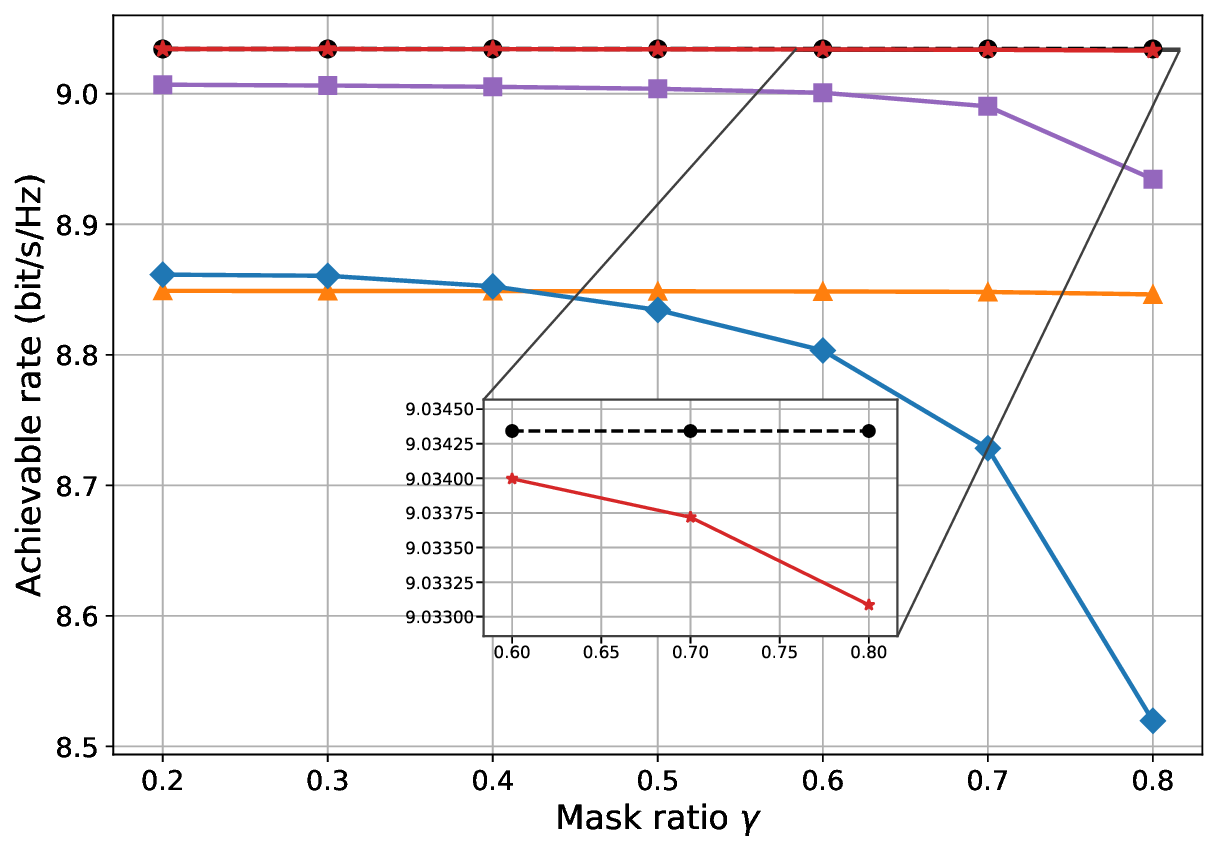}%
}\hspace{\colgap}%
\subfigure[$\mathrm{SNR}=10$ dB]{%
    \includegraphics[width=\w]{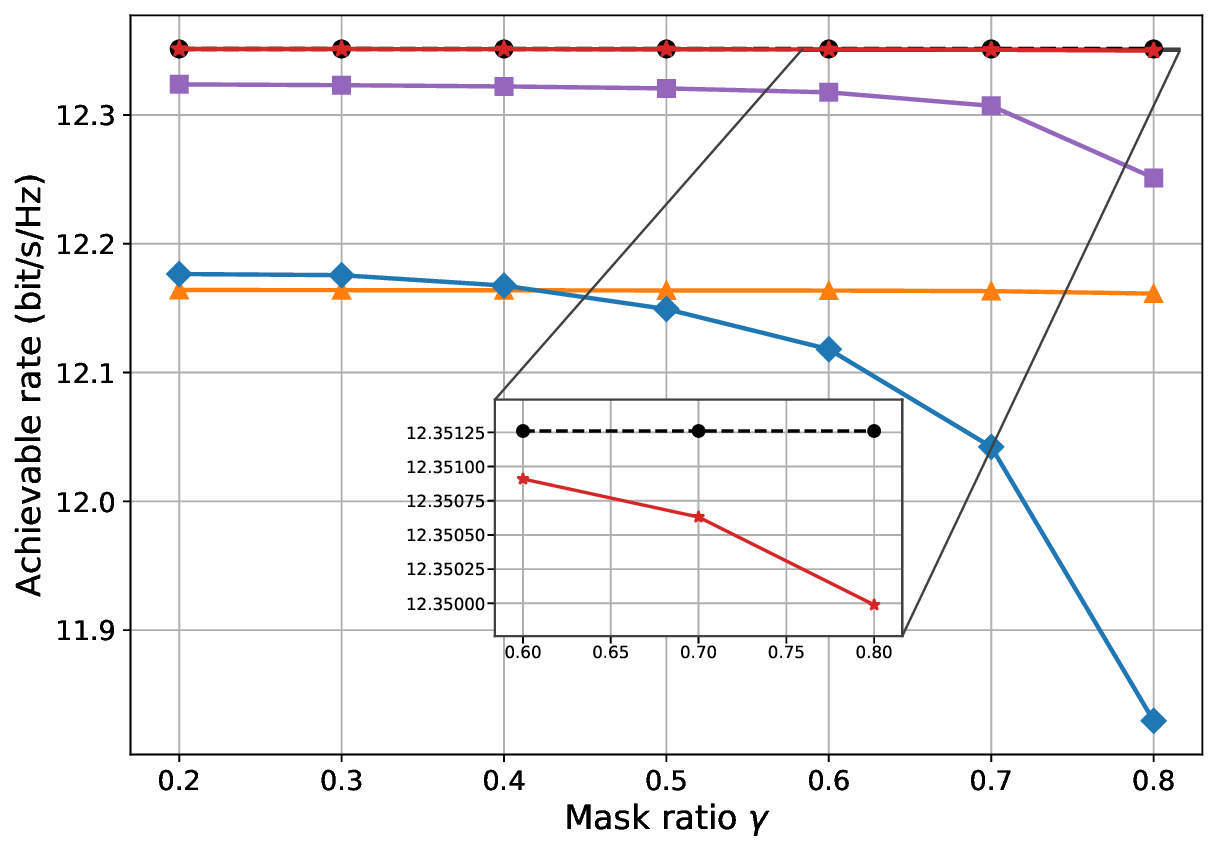}%
}

\caption{{\bl Achievable-rate comparison under antenna-domain random masking at different SNR levels.}}
\label{fig:achievable_rate_results}
\end{figure*}

\begin{figure*}[t]
    \centering
    \includegraphics[width=0.8\linewidth]{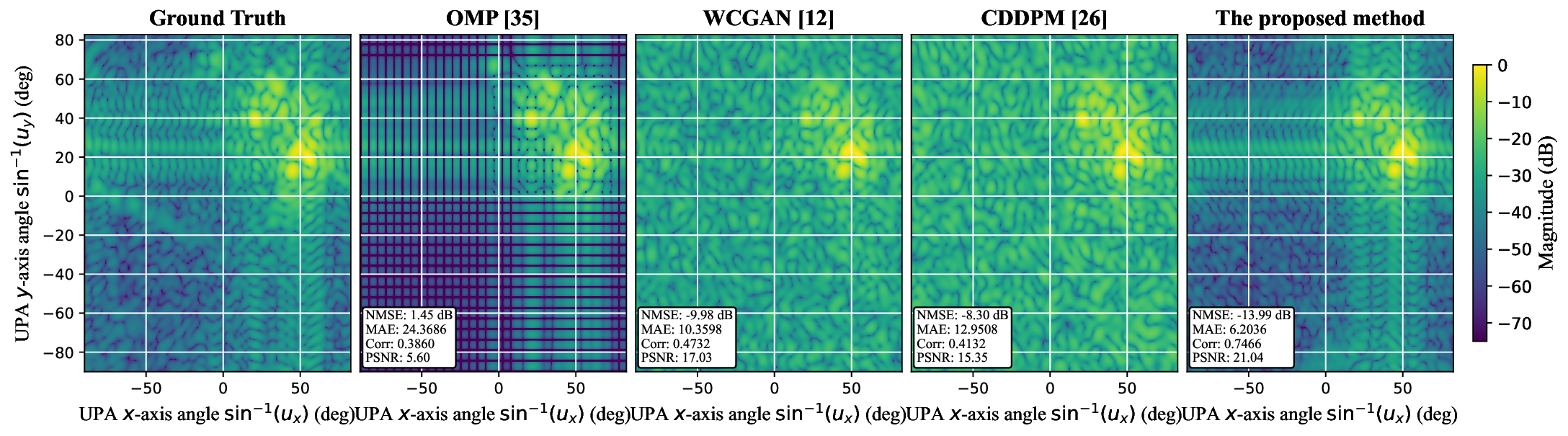}
\caption{{\bl A representative sample of the extrapolated channel in the angular domain under random masking ($\gamma=0.5$).}}
    \label{fig:multipath_results}
\end{figure*}

\subsection{Antenna-Domain Channel Extrapolation}

We first evaluate antenna-domain extrapolation performance under both random and structured mask schemes. Fig.~\ref{fig:antenna_results} illustrates the comparative results. Under both random and structured mask schemes, the proposed WGAN-enhanced CDDIM outperforms other baselines on NMSE, L1, and cosine distance, demonstrating robust extrapolation under partial observations. {\bl Notably, the WGAN-based baseline exhibits nearly invariant performance as the mask ratio $\gamma$ increases under random masking, since its one-shot adversarial completion mainly emphasizes distribution-level plausibility rather than explicit element-wise consistency with the observed CSI. In contrast, under structured masking, its performance degrades more clearly with $\gamma$ due to the lack of valid local priors inside large contiguous missing regions.} Conversely, while the CDDPM baseline is competitive when $\gamma \leq 0.4$, its accuracy collapses under heavy masking as it converges toward the zero-filling baseline. This reveals that standard diffusion models struggle to maintain structural integrity without XL-MIMO-specific priors, a failure mode our enhanced CDDIM successfully avoids.

Most notably, the proposed scheme maintains a substantial performance margin, outperforming all baseline algorithms by approximately 10-20 dB in terms of NMSE across the entire range of mask ratios and patterns. 
Moreover, regarding mask configurations, distinct behaviors are observed. 
For random masks, channel extrapolation performance exhibits a predictable degradation as $\gamma$ increases due to reduced known channel guidance. 
Conversely, it can be observed that diffusion-based methods typically exhibit a counter-intuitive non-monotonic pattern under structured masking, where reconstruction fidelity under heavier masking can paradoxically exceed that of moderate masking. 
Taking the proposed algorithm as an instance, the performance at $\gamma \in \{0.6, 0.7, 0.8\}$ surpasses that at $\gamma=0.5$. 
This phenomenon is driven by boundary-dilution effects as explained in~\cite{lugmayr2022repaint} that the perimeter-to-area ratio decreases as the missing block expands, reducing the proportion of error-prone edges. 
{\bl More specifically, the transition region between the observed and missing entries is more prone to reconstruction mismatch, whereas the interior of a large missing block is mainly governed by the learned global channel prior. 
Therefore, when the missing block becomes larger, the boundary-induced error is averaged over a larger missing area, making its contribution to the mean reconstruction error smaller. 
This explains why a slight non-monotonic trend can occur under structured masking, although the overall task remains more challenging as the available observations decrease.}

\begin{figure*}[t]
\centering

\newcommand{\figw}{0.25\linewidth}
\newcommand{\figgap}{-0.25em}

\subfigure[Random, frequency]{%
  \includegraphics[width=\figw]{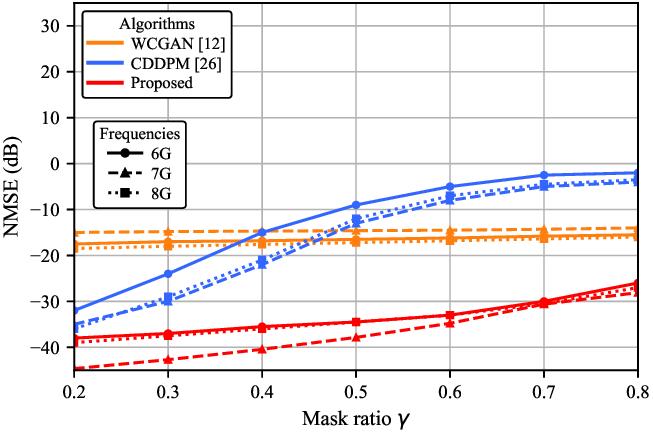}%
}\hspace{\figgap}%
\subfigure[Random, array/field]{%
  \includegraphics[width=\figw]{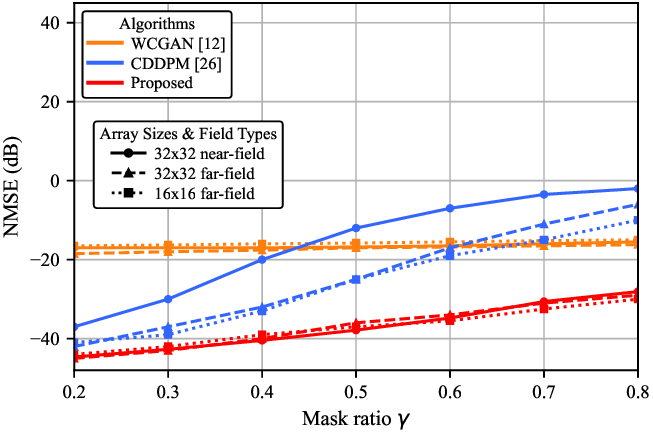}%
}\hspace{\figgap}%
\subfigure[Structured, frequency]{%
  \includegraphics[width=\figw]{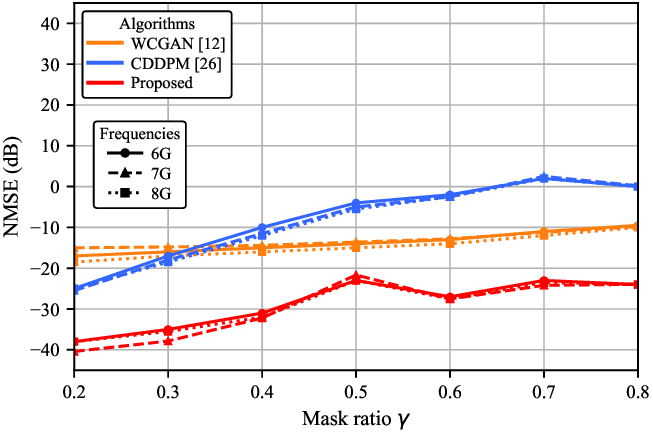}%
}\hspace{\figgap}%
\subfigure[Structured, array/field]{%
  \includegraphics[width=\figw]{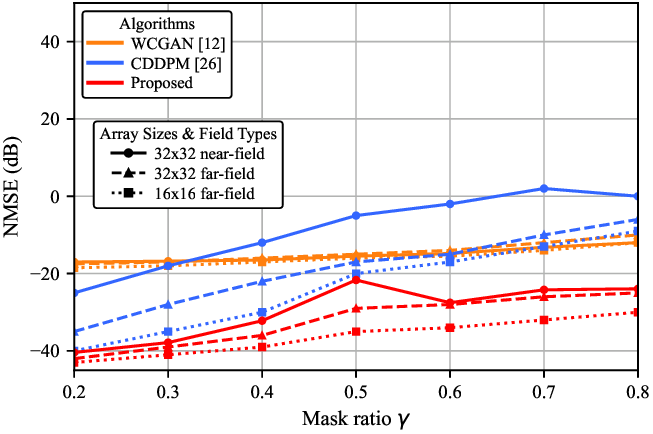}%
}

\vspace{-0.5em}
\caption{Model generalization evaluation under different mask schemes. 
(a) Random mask with training at 7\,GHz and testing at 6/7/8\,GHz. 
(b) Random mask with training/testing under $16{\times}16$ far-field, $32{\times}32$ near-field, and $32{\times}32$ far-field settings. 
(c) Structured mask with training at 7\,GHz and testing at 6/7/8\,GHz. 
(d) Structured mask with training/testing under different array sizes and field regimes.}
\label{fig:antenna_results_supplementary}
\end{figure*}

\newcommand{\imgwidth}{0.7\linewidth} 
\newcommand{\imgvspace}{-0.4em}       

\begin{figure}[t]
\centering
\subfigure[RePaint scheme comparison with random masks.]{%
    \includegraphics[width=\imgwidth]{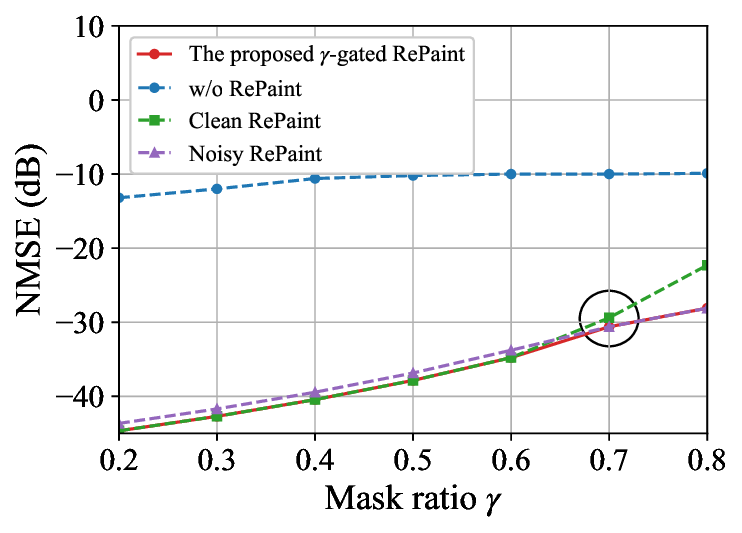}%
}\vspace{\imgvspace}

\subfigure[RePaint scheme comparison with structured masks.]{%
    \includegraphics[width=\imgwidth]{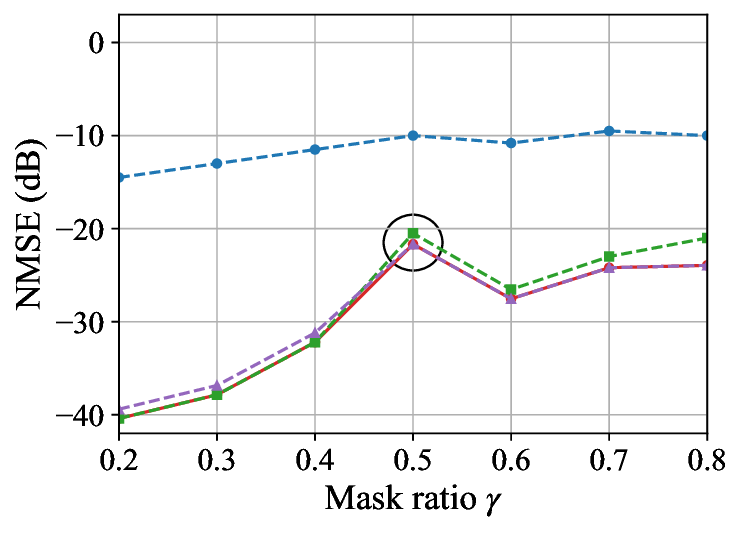}%
}

\caption{Ablation study on the proposed $\gamma$-gated RePaint scheme under both random mask and structured mask in antenna domain.}
\label{fig:RePaint}
\end{figure}

An ablation study on the proposed WGAN-guidance mechanism is also provided in Fig.~\ref{fig:antenna_results}. The proposed physics-aware CDDIM without WGAN supervision performs on par with, or close to, the proposed supervised version at low mask levels ($\gamma \leq 0.2$). As the mask ratio grows, the proposed WGAN-enhanced CDDIM exhibits increasingly superior performance, particularly under structured masks, indicating that WGAN supervision both expedites convergence and provides effective guidance toward the correct denoising direction.

{\bl We also calculate the achievable rate based on the extrapolated antenna-domain CSI to evaluate the end-to-end communication performance. Specifically, for each extrapolated channel $\hat{\mathbf h}$, we construct a maximum-ratio transmission (MRT) beamformer as $\mathbf w=\hat{\mathbf h}/\|\hat{\mathbf h}\|_2$, and then evaluate the achievable rate over the GT channel $\mathbf h$ by $R_{\mathrm{ach}}=\log_2\left(1+\rho_{tx}|\mathbf h^H\mathbf w|^2\right)$, where $\rho_{tx}$ denotes the transmit SNR. As shown in Fig.~\ref{fig:achievable_rate_results}, the proposed method almost achieves the perfect-CSI benchmark under different mask ratios and SNR levels. In contrast, OMP and CDDPM suffer noticeable rate losses under severe masking, while WCGAN achieves stable but lower rates. Moreover, the performance advantage of the proposed method is consistently observed at all tested SNR levels. At low SNR, the rate gap among different methods is relatively small because the noise-limited regime weakens the impact of CSI errors, whereas at higher SNR, the degradation caused by inaccurate CSI becomes more evident. The proposed method remains close to the perfect-CSI benchmark, indicating that its improved channel extrapolation accuracy is effectively translated into beamforming gains.}

{\bl To further evaluate the angular-domain reconstruction accuracy, we consider $\gamma=0.5$ under the random mask scheme and visualize a representative sample in the angular domain. 
As shown in Fig.~\ref{fig:multipath_results}, the GT angular-domain channel exhibits a dominant energy-concentrated region together with several weaker angular components. 
The OMP-based method suffers from evident grid-like artifacts, while WCGAN and CDDPM can recover the main energy region but tend to produce smoother or less accurate weak-component structures. 
In comparison, the proposed method better preserves the dominant angular energy distribution and reconstructs the surrounding weaker components more consistently with the GT. 
This visual observation is also consistent with the quantitative results, where the proposed method achieves the lowest NMSE and MAE, as well as the highest correlation and PSNR, indicating improved angular-domain reconstruction fidelity.}

For validating the generalization performance of our approach, we also conduct experiments across various central carrier frequencies and antenna sizes (with field regimes). Specifically, we compare WGAN, CDDPM, and the proposed WGAN-enhanced CDDIM at central frequencies of 6\,GHz, 7\,GHz, and 8\,GHz, and across antenna configurations and regimes (${16\times 16}$ far-field, ${32\times 32}$ near-field, and ${32\times 32}$ far-field). The results are summarized in Fig.~\ref{fig:antenna_results_supplementary}. 
Figs.~\ref{fig:antenna_results_supplementary}(a) and~\ref{fig:antenna_results_supplementary}(c) show the performance under random and structured masks across different central carrier frequencies, respectively. 
The proposed WGAN-enhanced CDDIM consistently achieves the lowest NMSE, demonstrating robust frequency generalization. 
Figs.~\ref{fig:antenna_results_supplementary}(b) and~\ref{fig:antenna_results_supplementary}(d) further compare different array sizes and near-/far-field configurations under random and structured masks. 
The proposed method remains superior in these settings, indicating good scalability with respect to array size and radiation-field regimes.

Finally, to validate the effectiveness of the proposed $\gamma$-gated RePaint scheme, we also conduct an ablation study in the antenna domain under both random mask and structured mask settings, comparing no RePaint, clean RePaint, and noisy RePaint, as shown in Fig.~\ref{fig:RePaint}. Removing the RePaint step weakens the consistency with the observed entries during reverse sampling, leading to a significant performance drop of 14--27 dB. This underscores the necessity of injecting known-channel priors during reverse diffusion. Moreover, a distinct trade-off can be observed that clean RePaint performs better at low mask ratios ($\gamma<0.7$ under random mask and $\gamma<0.5$ under structured mask), whereas noisy RePaint excels at high mask ratios, and their performance gap can be up to $6$ dB. By adaptively combining the two strategies, we set the gating thresholds to $\gamma_{\text{thre}}=0.7$ for random masks and $\gamma_{\text{thre}}=0.5$ for structured masks in our final implementation. The proposed $\gamma$-gated RePaint inherits the advantages of both and attains the best performance across all mask ratios.





\begin{figure*}[t]
\centering

\newcommand{\figw}{0.3\linewidth}
\newcommand{\colgap}{-0.8em}
\newcommand{\rowgap}{-0.25em}

\subfigure[NMSE, frequency domain.]{
    \includegraphics[width=\figw]{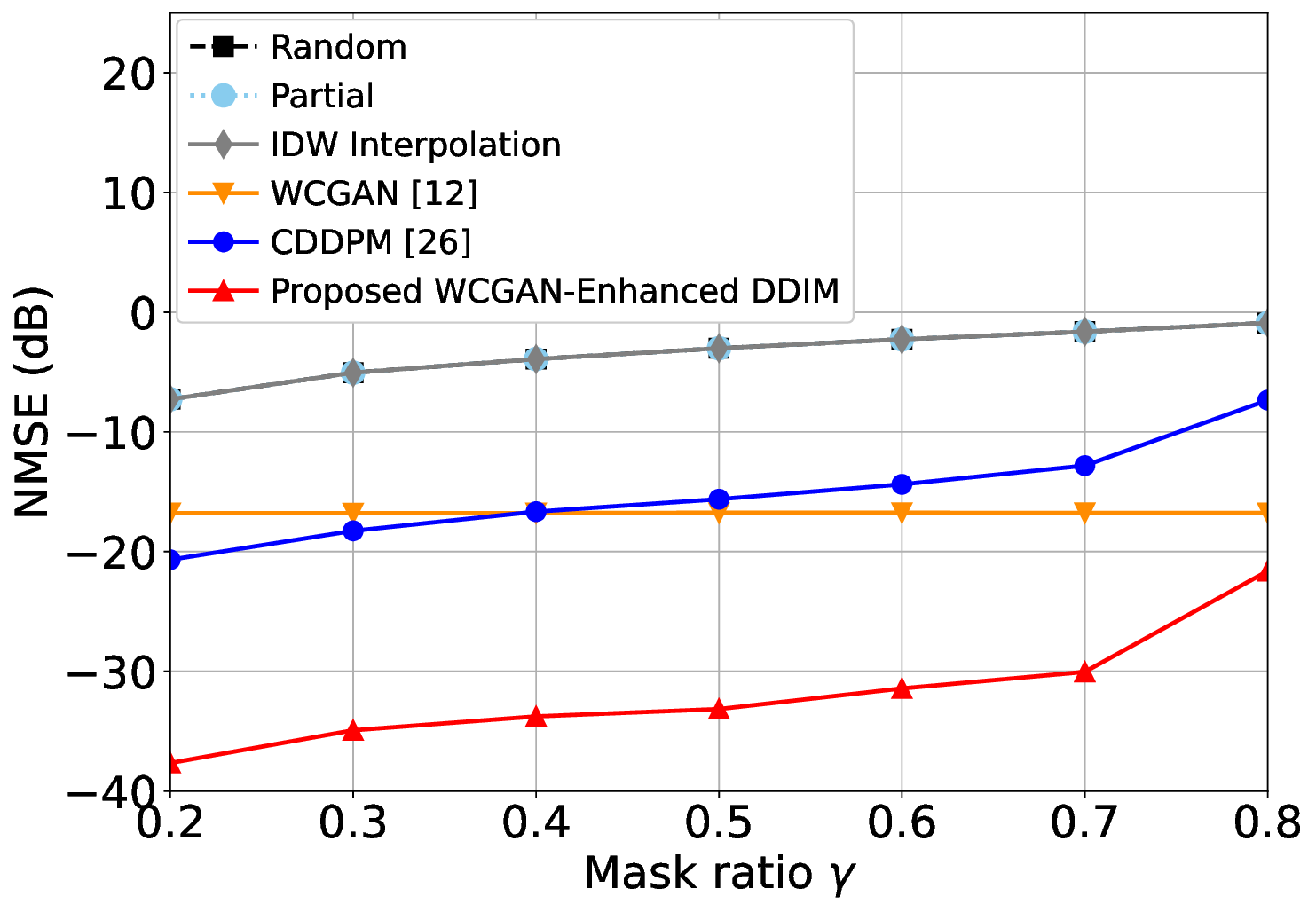}
}
\hspace{\colgap}
\subfigure[$L_1$ distance, frequency domain.]{
    \includegraphics[width=\figw]{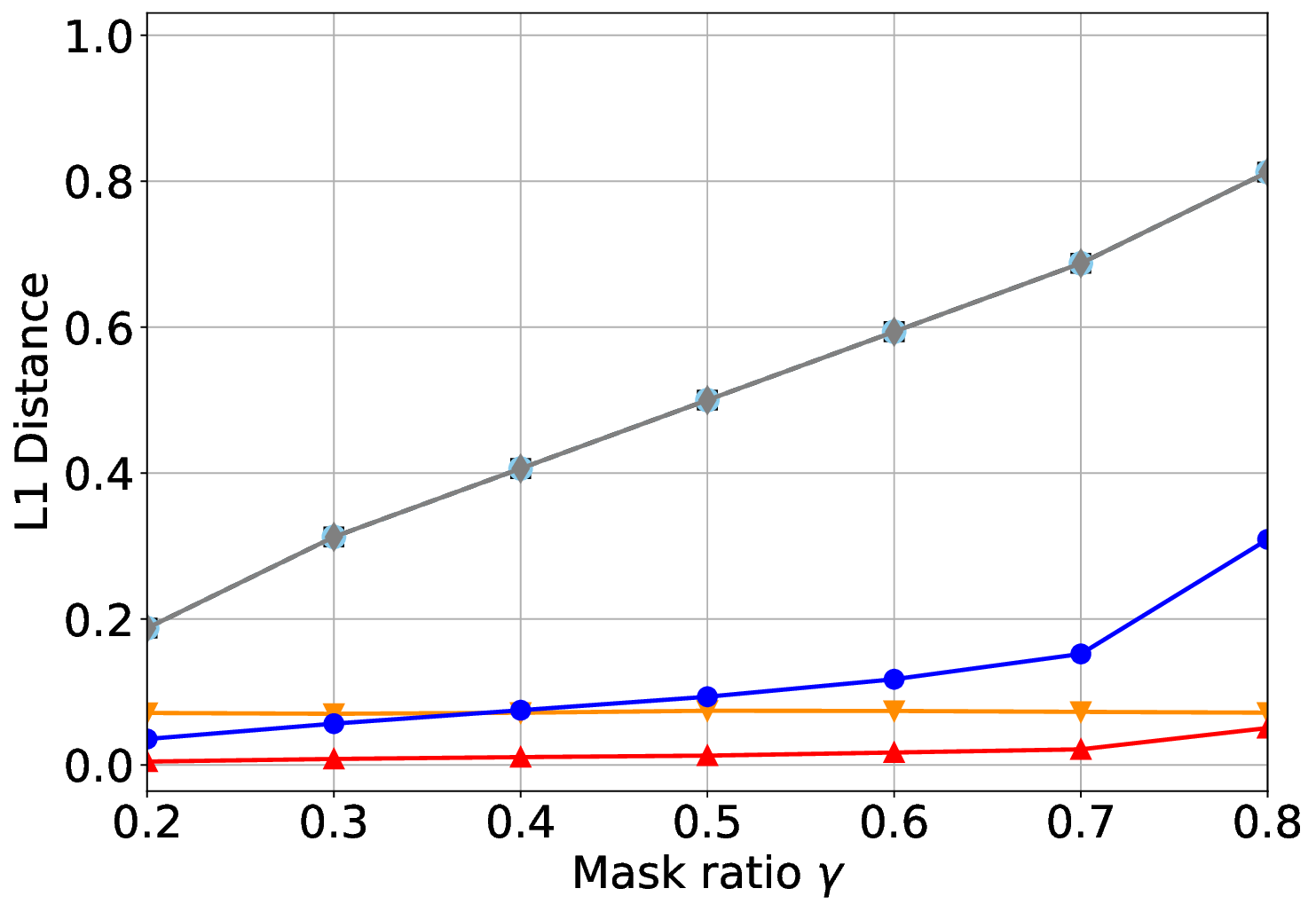}
}
\hspace{\colgap}
\subfigure[Cosine distance, frequency domain.]{
    \includegraphics[width=\figw]{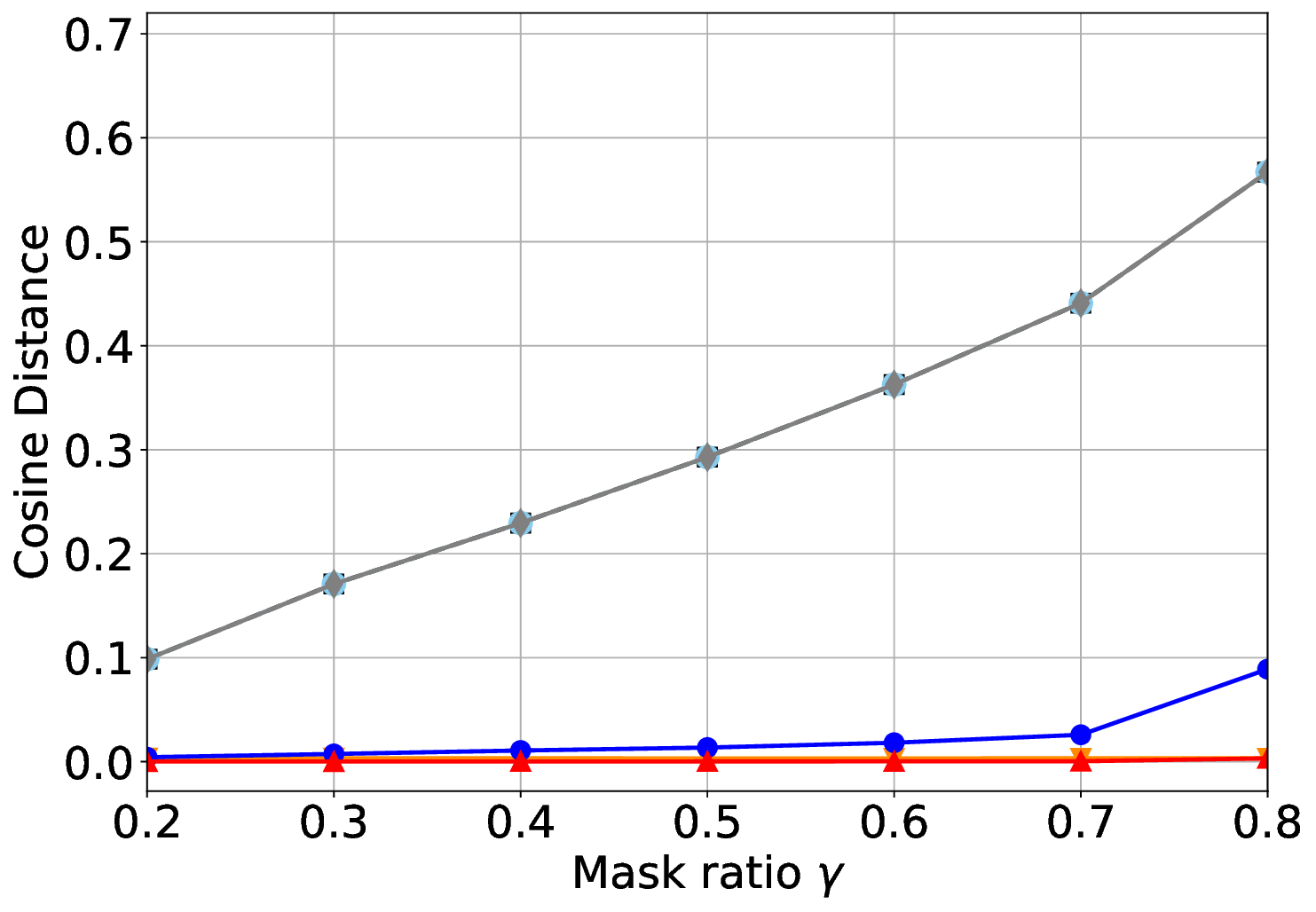}
}

\vspace{\rowgap}

\subfigure[NMSE, spatial domain.]{
    \includegraphics[width=\figw]{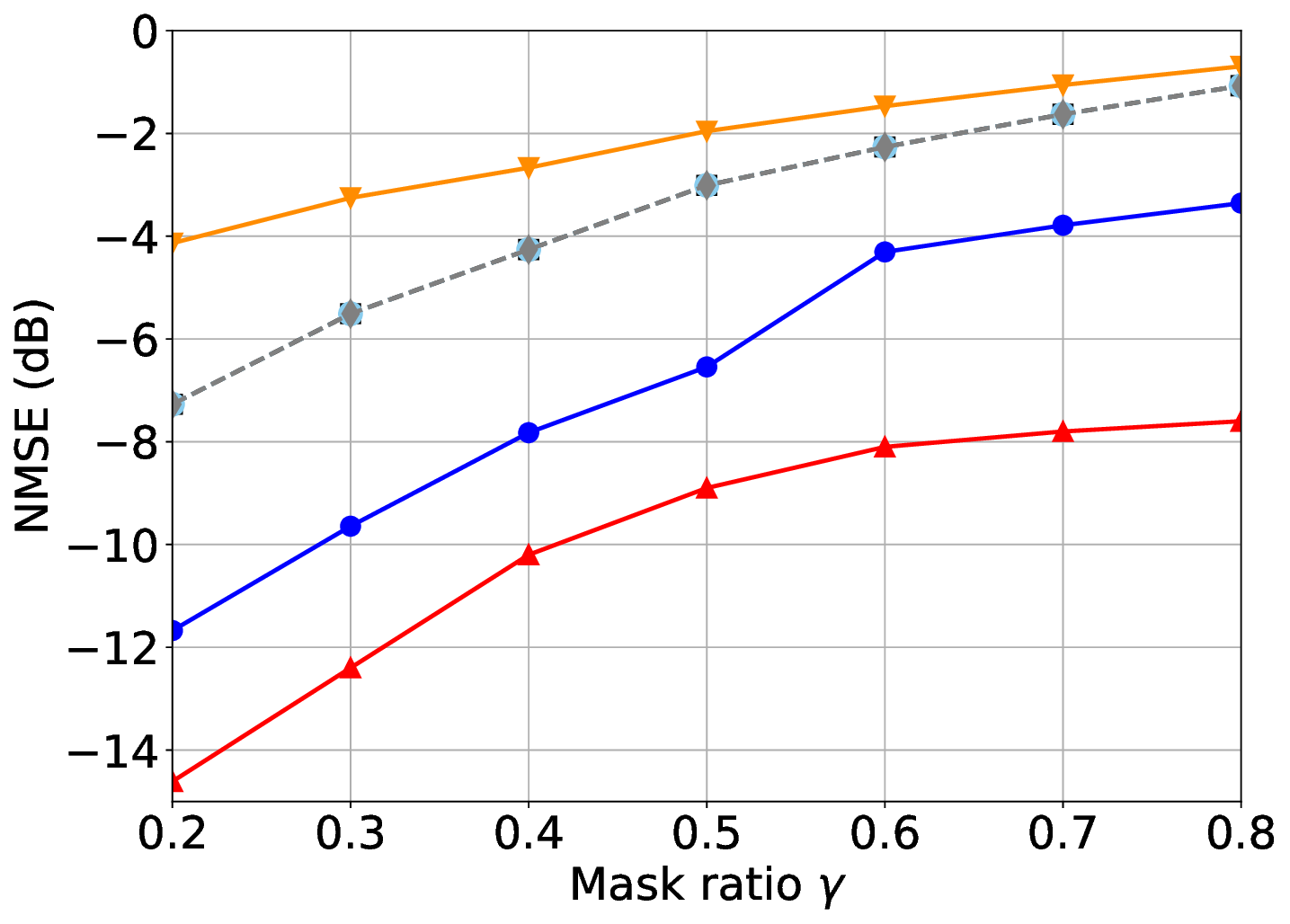}
}
\hspace{\colgap}
\subfigure[$L_1$ distance, spatial domain.]{
    \includegraphics[width=\figw]{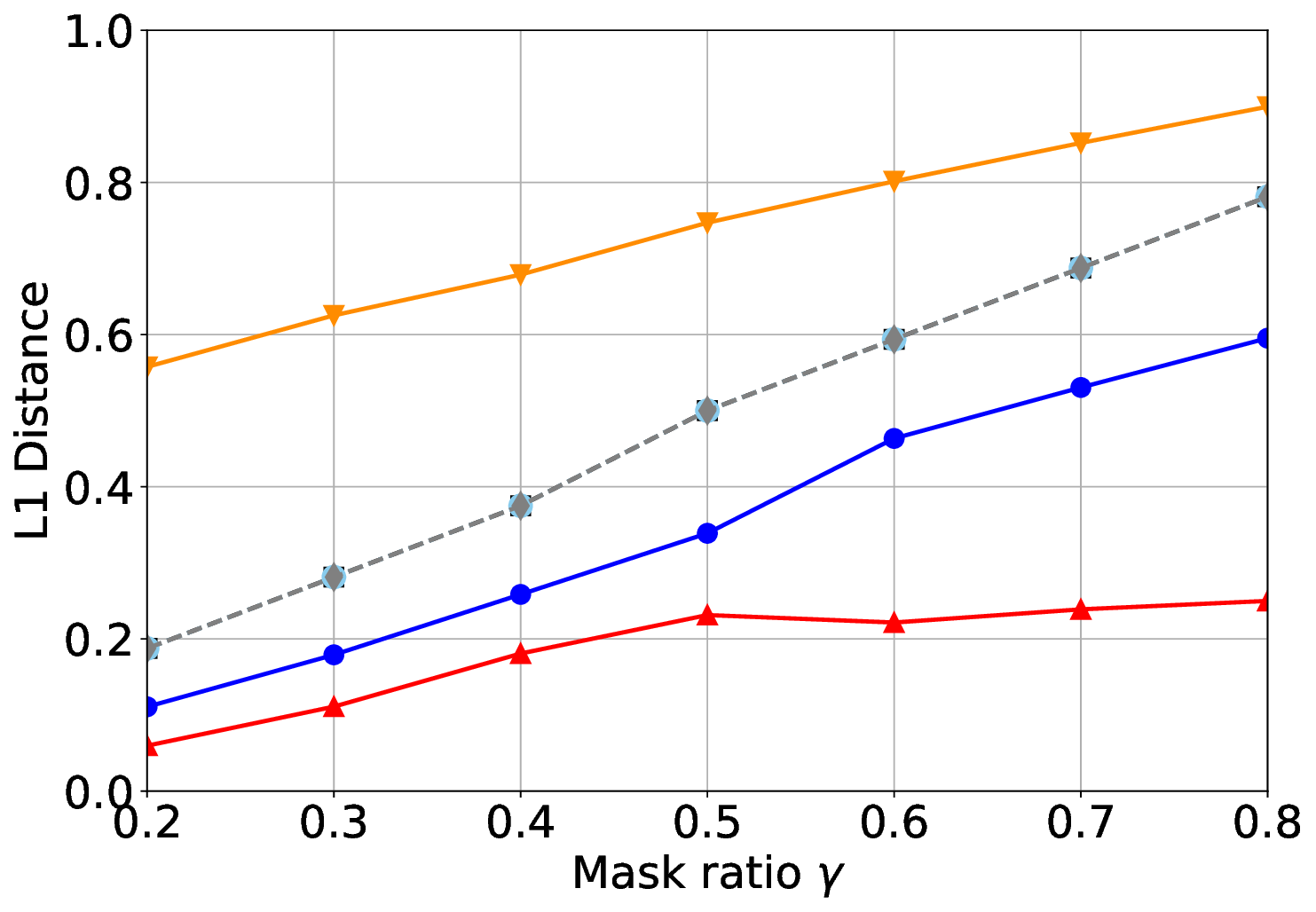}
}
\hspace{\colgap}
\subfigure[Cosine distance, spatial domain.]{
    \includegraphics[width=\figw]{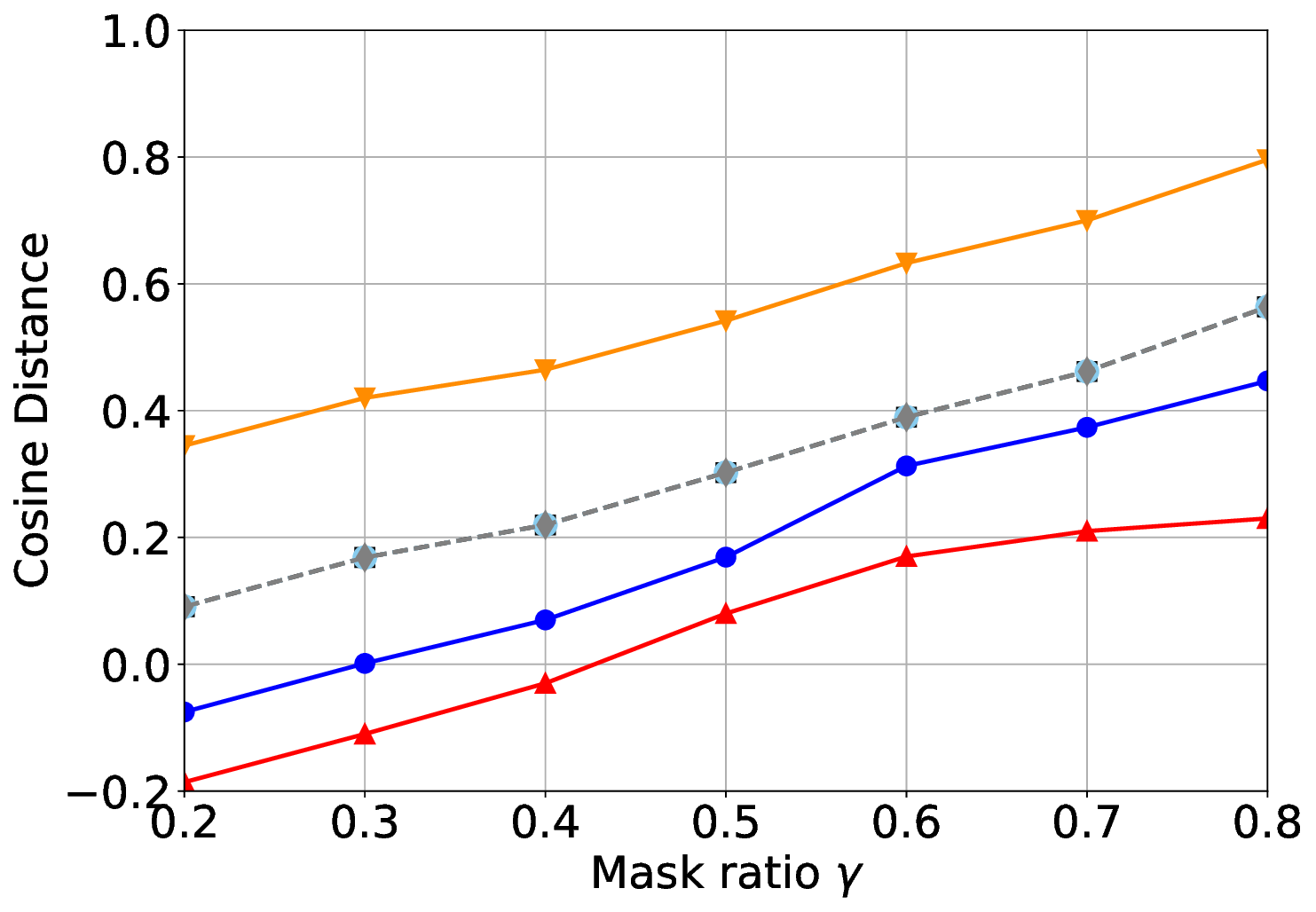}
}

\caption{{\bl Channel extrapolation performance in the frequency and spatial domains. Frequency-domain extrapolation evaluated by (a) NMSE, (b) $L_1$ distance, and (c) cosine distance. 
Spatial-domain extrapolation evaluated by (d) NMSE, (e) $L_1$ distance, and (f) cosine distance.}}
\label{fig:results_fre_and_spatial}
\end{figure*}

\subsection{Frequency- and Spatial-Domain Channel Extrapolation}

We further extend the evaluation to the high-dimensional 3D frequency and spatial domains, as summarized in Fig.~\ref{fig:results_fre_and_spatial}. 
{\bl For frequency-domain extrapolation, Figs.~\ref{fig:results_fre_and_spatial}(a)--(c) show that the proposed WGAN-enhanced CDDIM consistently achieves the best performance in terms of NMSE, $L_1$ distance, and cosine distance. 
For spatial-domain extrapolation, Figs.~\ref{fig:results_fre_and_spatial}(d)--(f) show a similar trend. 
These results validate the robustness of the proposed method in both frequency- and spatial-domain extrapolation, while WCGAN and CDDPM exhibit more evident degradation as the mask ratio increases. The observed relatively higher NMSE in the spatial-domain task is mainly due to the
faster spatial correlation decay.}

\begin{figure*}[t]
    \centering
    \includegraphics[width=0.82\linewidth]{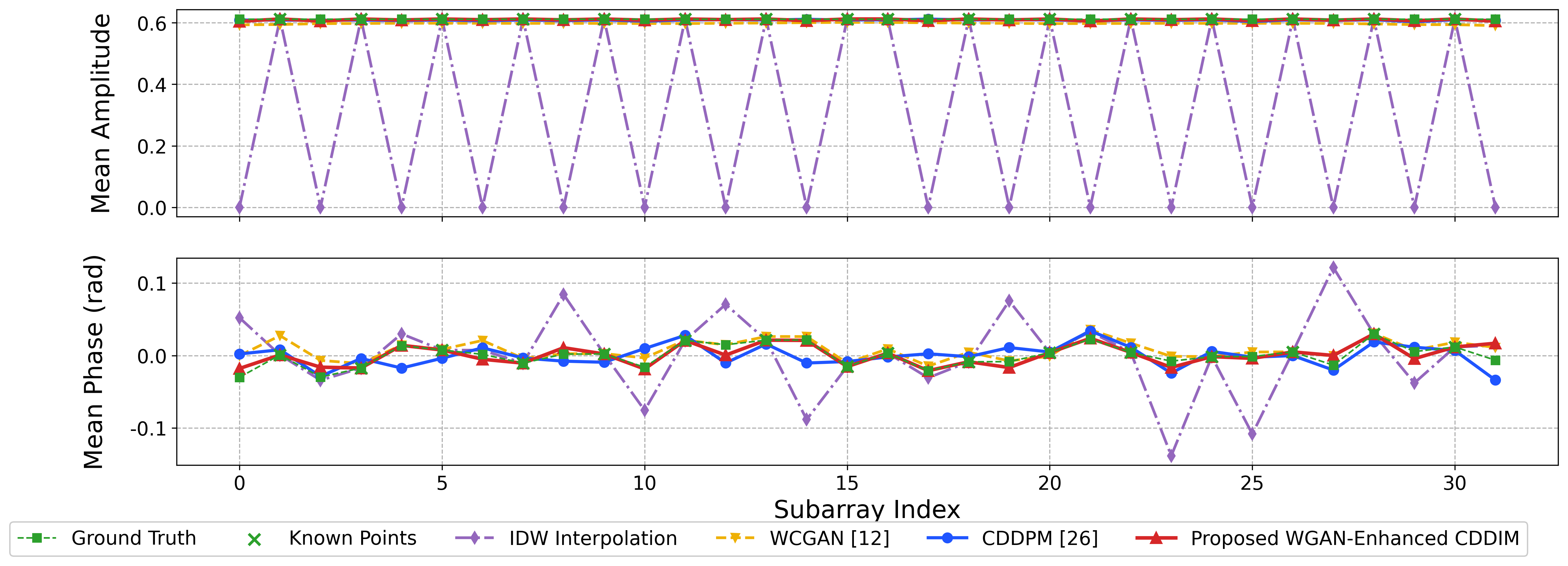}
    \caption{Comparison of mean amplitude and phase utilizing different extrapolation methods ($\gamma = 0.5$).}
    \label{fig:results_frequency_analysis}
\end{figure*}

To provide an in-depth comparison of channel extrapolation quality, we select a representative sample in the frequency domain at $\gamma = 0.5$ and visualize the mean amplitude and phase over subcarriers in Fig.~\ref{fig:results_frequency_analysis}.
 It can be observed that IDW interpolation fails to recover valid magnitudes, yielding near-zero amplitudes in masked bands. WCGAN improves over IDW but still exhibits a slight amplitude bias, while CDDPM reduces the magnitude underestimation but shows a larger phase bias. In contrast, our method closely follows the GT across all subcarriers. For phase, IDW incurs large deviations, WCGAN remains moderately biased, and CDDPM’s bias is more evident, whereas our model maintains phase continuity with minimal error, preserving the underlying angular structure under sparse observations.

\begin{table*}[t]
\centering
\caption{{\bl Complexity and representative accuracy comparison under different sampling steps for 2D and 3D extrapolation tasks.}}
\label{tab:complexity}
\footnotesize
\renewcommand{\arraystretch}{0.95}
\setlength{\tabcolsep}{3.8pt}
\begin{tabular}{@{}lcccc ccc@{}}
\toprule
\multirow{2}{*}{Model} 
& \multirow{2}{*}{Steps} 
& \multicolumn{3}{c}{2D antenna domain} 
& \multicolumn{3}{c}{3D frequency/spatial domain} \\
\cmidrule(lr){3-5} \cmidrule(lr){6-8}
& & Latency (ms) & GFLOPs & Avg. NMSE (dB)$^{\ddagger}$ 
  & Latency (ms) & GFLOPs & Avg. NMSE (dB)$^{\dagger}$ \\
\midrule
WCGAN~\cite{balevi2021wcgan} 
& 1 
& 2.03 & 3.18 & -14.06 
& 9.15 & 152.54 & -16.93 \\

CDDPM~\cite{zhang2025cddpm} 
& 1000 
& 112.34 & 107.15 & -15.33 
& 873.48 & 9827.65 & -15.11 \\
\midrule
Proposed method 
& 5   
& 1.91  & 0.54 & -23.66 
& 7.80 & 49.13 & -18.59 \\

Proposed method 
& 10  
& 2.79  & 1.07 & -27.83 
& 11.67 & 98.26 & -21.35 \\

Proposed method 
& 20  
& 4.36  & 2.14 & -31.35 
& 20.07 & 196.52 & -25.41 \\

Proposed method 
& 50  
& 9.80 & 5.36 & -35.33 
& 47.24 & 491.32 & -28.96 \\

\textbf{Proposed method}
& \textbf{100 (adopted)}
& \textbf{19.87} & \textbf{10.71} & \textbf{-37.18}
& \textbf{91.12} & \textbf{982.64} & \textbf{-30.23} \\

Proposed method 
& 200  
& 38.67 & 21.43 & -38.15
& 175.51 & 1965.29 & -30.83 \\
\bottomrule
\end{tabular}

\vspace{0.3em}
\begin{minipage}{0.98\linewidth}
\footnotesize
\textit{Note:}$^{\ddagger}$ The 2D Avg. NMSE is evaluated under the antenna-domain random mask setting. $^{\dagger}$ For the 3D domain, the reported Avg. NMSE is evaluated on the frequency-domain extrapolation results.
\end{minipage}
\end{table*}

\subsection{Computational Complexity Analysis}
{\bl
Finally, we compare the computational cost, inference latency, and representative extrapolation accuracy of different methods in both 2D and 3D tasks, as summarized in Table~\ref{tab:complexity}. 
The estimated GFLOPs are computed from the main inference network and serve as a reference for theoretical arithmetic complexity. 
For diffusion-based methods, different reverse sampling steps are also evaluated to characterize the accuracy-complexity trade-off.

At small sampling steps, the proposed method achieves higher accuracy with comparable or even lower latency than WCGAN. 
For the 2D antenna-domain task, the proposed method with $5$ sampling steps reduces the latency from $2.03$ ms to $1.91$ ms and improves the average NMSE from $-14.06$ dB to $-23.66$ dB compared with WCGAN. 
For the 3D frequency-domain task, it also reduces the latency from $9.15$ ms to $7.80$ ms while improving the average NMSE from $-16.93$ dB to $-18.59$ dB. 
When the number of steps increases to $10$, the latency remains close to that of WCGAN, while the average NMSE further improves to $-21.35$ dB.

Compared with the conventional CDDPM baseline, the proposed method substantially reduces inference latency and computational cost while achieving higher reconstruction accuracy. 
The $100$-step setting is adopted as the default configuration because it provides a good balance between accuracy and latency.
Further increasing the sampling steps from $100$ to $200$ yields only marginal NMSE gains but nearly doubles the inference latency. 
}

\section{Conclusion}
\label{sec:conclusion}
In this paper, we proposed an efficient multi-domain near-field channel extrapolation framework for XL-MIMO systems. To overcome the challenges of high-dimensional CSI acquisition and spherical wavefront propagation, a physics-aware CDDIM backbone was developed to extrapolate channels across the antenna, frequency, and spatial domains. By incorporating position-embedded patch tokenization and mask-guided multi-head attention, the proposed model effectively captures position-dependent near-field channel correlations. Furthermore, WGAN-guided adversarial supervision and a RePaint-style refinement scheme were introduced to improve the fidelity of the reverse sampling process, especially under severe masking conditions. Extensive experiments demonstrated that the proposed framework achieves superior extrapolation accuracy and robust generalization across diverse domains, frequencies, array configurations, and mask patterns. In addition, the use of non-Markovian DDIM sampling substantially reduces inference latency and computational complexity, making the proposed framework a practical solution for efficient near-field XL-MIMO channel acquisition. {\bl Extending the proposed framework to multi-user and multi-antenna scenarios is also an important direction for future work, where the spatial sampling dimension can be generalized to a location-aware user dimension, and an additional UE-antenna dimension can be incorporated through UE-side positional encoding and generalized patch embedding.}

\appendices
\section{Derivation of the posterior distribution in \eqref{eq:posterior-bayes-gauss}}
\label{app:posterior-derivation}
By Bayes' rule and Gaussian conjugacy for the forward diffusion process, the posterior in~\eqref{eq:posterior-bayes-gauss} can be expanded as
{\begingroup
\allowdisplaybreaks[4]
\setlength{\jot}{1pt}
\setlength{\arraycolsep}{1pt}
\footnotesize

\begin{align}
q(&\mathbf{H}_{t-1} \mid \mathbf{H}_t, \mathbf{H}_0) \notag \\
=& \frac{q(\mathbf{H}_t \mid \mathbf{H}_{t-1}) \, q(\mathbf{H}_{t-1} \mid \mathbf{H}_0)}{q(\mathbf{H}_t \mid \mathbf{H}_0)} \notag \\
=& \frac{\mathcal{CN}\!\left( \mathbf{H}_t; \sqrt{\alpha_t}\,\mathbf{H}_{t-1}, (1-\alpha_t)\mathbf{I} \right)
       \mathcal{CN}\!\left( \mathbf{H}_{t-1}; \sqrt{\bar{\alpha}_{t-1}}\,\mathbf{H}_0, (1-\bar{\alpha}_{t-1})\mathbf{I} \right)}
     {\mathcal{CN}\!\left( \mathbf{H}_t; \sqrt{\bar{\alpha}_t}\,\mathbf{H}_0, (1-\bar{\alpha}_t)\mathbf{I} \right)} \notag \\
\propto & \exp\!\Bigg\{ -\frac{1}{2} \bigg[
\frac{\|\mathbf{H}_t-\sqrt{\alpha_t}\mathbf{H}_{t-1}\|_F^2}{1-\alpha_t}
+ \frac{\|\mathbf{H}_{t-1}-\sqrt{\bar{\alpha}_{t-1}}\mathbf{H}_0\|_F^2}{1-\bar{\alpha}_{t-1}}\notag \\
&- \frac{\|\mathbf{H}_t-\sqrt{\bar{\alpha}_t}\mathbf{H}_0\|_F^2}{1-\bar{\alpha}_t}
\bigg] \Bigg\} \notag \\
\propto & \exp\!\Bigg\{ -\frac{1}{2} \bigg[
\Big( \frac{\alpha_t}{1-\alpha_t} + \frac{1}{1-\bar{\alpha}_{t-1}} \Big) \|\mathbf{H}_{t-1}\|_F^2
- 2\,\Re\!\Big\langle \frac{\sqrt{\alpha_t}}{1-\alpha_t}\mathbf{H}_t \notag \\
&+ \frac{\sqrt{\bar{\alpha}_{t-1}}}{1-\bar{\alpha}_{t-1}} \mathbf{H}_0,\ \mathbf{H}_{t-1} \Big\rangle_F
\bigg] \Bigg\} \notag \\
=& \mathcal{CN}\!\bigl(
\mathbf{H}_{t-1};\,
\frac{\sqrt{\bar{\alpha}_{t-1}}(1-\alpha_t)}{1-\bar{\alpha}_t}\,\mathbf{H}_0
+ \frac{\sqrt{\alpha_t}(1-\bar{\alpha}_{t-1})}{1-\bar{\alpha}_t}\,\mathbf{H}_t,\notag\\
&\qquad \frac{1-\bar{\alpha}_{t-1}}{1-\bar{\alpha}_t}(1-\alpha_t)\,\mathbf{I}
\bigr) \notag\\
=& \mathcal{CN}\!\bigl( \mathbf{H}_{t-1};
\underbrace{\tfrac{1}{\sqrt{\alpha_t}}\bigl( \mathbf{H}_t - \tfrac{1-\alpha_t}{\sqrt{1-\bar{\alpha}_t}}\boldsymbol{\epsilon}_t \bigr)}_{\boldsymbol{\mu}(\mathbf{H}_t,\mathbf{H}_0,t)},
\underbrace{\tfrac{\beta_t (1-\bar{\alpha}_{t-1})}{1-\bar{\alpha}_t}}_{\boldsymbol{\Sigma}_t}\, \mathbf{I} \bigr).
\end{align}
\endgroup}

{\bl
\section{Derivation of the Deterministic Non-Markovian Reverse Sampling Trajectory in~\eqref{eq:reverse_sampling}}
\label{app:ddim-trajectory}

We derive the deterministic reverse sampling trajectory in~\eqref{eq:reverse_sampling}. 
From the forward marginal in~\eqref{eq:noisy_channel}, the clean channel can be estimated at reverse step $t_2$ using the predicted noise
\begin{equation}
\hat{\boldsymbol{\epsilon}}_{t_2}
=
\mathcal G_{\theta}
(\mathbf H_{t_2},\mathbf H_{\mathrm{partial}},\mathbf M,t_2),
\end{equation}
which gives the one-step clean-channel estimate $\hat{\mathbf H}_{0|t_2}$ as defined in~\eqref{eq:One-step DDIM}.

For a skipped reverse transition from $t_2$ to $t_1$ with $t_1<t_2$, the deterministic non-Markovian trajectory combines the estimated clean component and the reused noise direction. 
This leads to
\begin{equation}
\mathbf H_{t_1|t_2}
=
\sqrt{\bar{\alpha}_{t_1}}\hat{\mathbf H}_{0|t_2}
+
\sqrt{1-\bar{\alpha}_{t_1}}\hat{\boldsymbol{\epsilon}}_{t_2}.
\label{eq:app-ddim-construct}
\end{equation}
The coefficient of $\hat{\boldsymbol{\epsilon}}_{t_2}$ is selected to match the scheduled noise level at time $t_1$.

Substituting the one-step estimate $\hat{\mathbf H}_{0|t_2}$ from~\eqref{eq:One-step DDIM} into~\eqref{eq:app-ddim-construct}, we can obtain
\begin{equation}
\begin{aligned}
\mathbf H_{t_1|t_2}
&=
\sqrt{\frac{\bar{\alpha}_{t_1}}{\bar{\alpha}_{t_2}}}
\mathbf H_{t_2}
\\
&\quad+
\left(
\sqrt{1-\bar{\alpha}_{t_1}}
-
\sqrt{\frac{(1-\bar{\alpha}_{t_2})\bar{\alpha}_{t_1}}{\bar{\alpha}_{t_2}}}
\right)
\hat{\boldsymbol{\epsilon}}_{t_2}.
\end{aligned}
\label{eq:app-ddim-expanded}
\end{equation}
Finally, replacing $\hat{\boldsymbol{\epsilon}}_{t_2}$ with the network output 
$\mathcal G_{\theta}(\mathbf H_{t_2},\mathbf H_{\mathrm{partial}},\mathbf M,t_2)$ gives~\eqref{eq:reverse_sampling}.
}

\bibliographystyle{IEEEtran}
\bibliography{ref-short}

\vspace{-1mm}

\vspace{12pt}
\color{red}

\end{document}